\documentclass[aps,preprint,amsmath,amssymb,floatfix]{revtex4-1}
\usepackage{graphicx,morefloats}
\usepackage[caption=false]{subfig}
\usepackage{color}

\begin{document}
\hyphenation{va-ni-sh-ing}

\begin{center}
{\large\bf
High Temperature Superconductivity in Iron Pnictides and Chalcogenides
}
\\[0.5cm]

%% Notice placement of commas and superscripts and use of &
%% in the author list

Qimiao Si$^{1}$, Rong Yu$^{2,3}$
\& Elihu Abrahams$^{4}$ \\

{\em $^1$Department of Physics and Astronomy, Rice University,
Houston, TX 77005, USA}\\

{\em $^2$Department of Physics and Beijing Key Laboratory of
Opto-electronic Functional Materials and Micro-nano Devices,
Renmin University, Beijing 100872, China}\\

{\em $^3$Department of Physics and Astronomy, Collaborative Innovation
Center of Advanced Microstructures, Shanghai Jiaotong University,
Shanghai 200240, China}

{\em $^4$Department of Physics and Astronomy, University of California Los Angeles,
Los Angeles, CA 90095, USA}\\

%%\maketitle

%%\begin{abstract}
%%\vspace{0.5cm}
\end{center}

\vspace{0.5cm}
{\bf
Superconductivity develops in metals upon the formation of a coherent macroscopic 
quantum state of electron pairs. Iron pnictides and chalcogenides are materials 
that have high superconducting transition temperatures. In this Review, 
we describe the advances in the field that have led to higher superconducting transition temperatures
in iron-based superconductors and the wide range of materials that form them. 
We summarize both the essential aspects of the normal state 
and the mechanism for superconductivity.  We emphasize the degree of electron-electron correlations 
and their manifestation in properties of the normal state. We examine the nature of magnetism, 
analyse its role in driving the electronic nematicity, and discuss quantum criticality 
at the border of magnetism in the phase diagram. Finally, we review the amplitude 
and structure of the superconducting pairing, and survey the potential settings 
for optimizing superconductivity.
}
%%\end{abstract}

%%\newpage

%\vskip 2 cm
\vskip 1 cm

In early 2008, the discovery of superconductivity with a transition temperature ($T_c$) of $26$ K
in an iron pnictide compound took the condensed matter and materials physics community
by surprise\cite{Hos2008}. It raised the prospect for high-temperature superconductivity
beyond the copper-based materials, the only materials known up to then having a $T_c$ higher than
$40$ K.\cite{Bednorz-Mueller1986}
While the transition temperature in the iron pnictides
was quickly raised to 56 K in a matter of a few months\cite{ChenRen2008},
that record did not change
for several years. However, recent developments in the iron
chalcogenides\cite{Xue.2012, SLHe2013,Shen.2014,YWang2015,Jia.2014}
have given renewed hope for even higher transition temperatures.
Meanwhile, considerable progress has been
made on the understanding of the microscopic physics of these iron-based superconductors (FeSCs).

 \begin{figure}[!ht]
\begin{center}
\includegraphics[
%angle=270,
angle=0,
width=0.9 \textwidth]{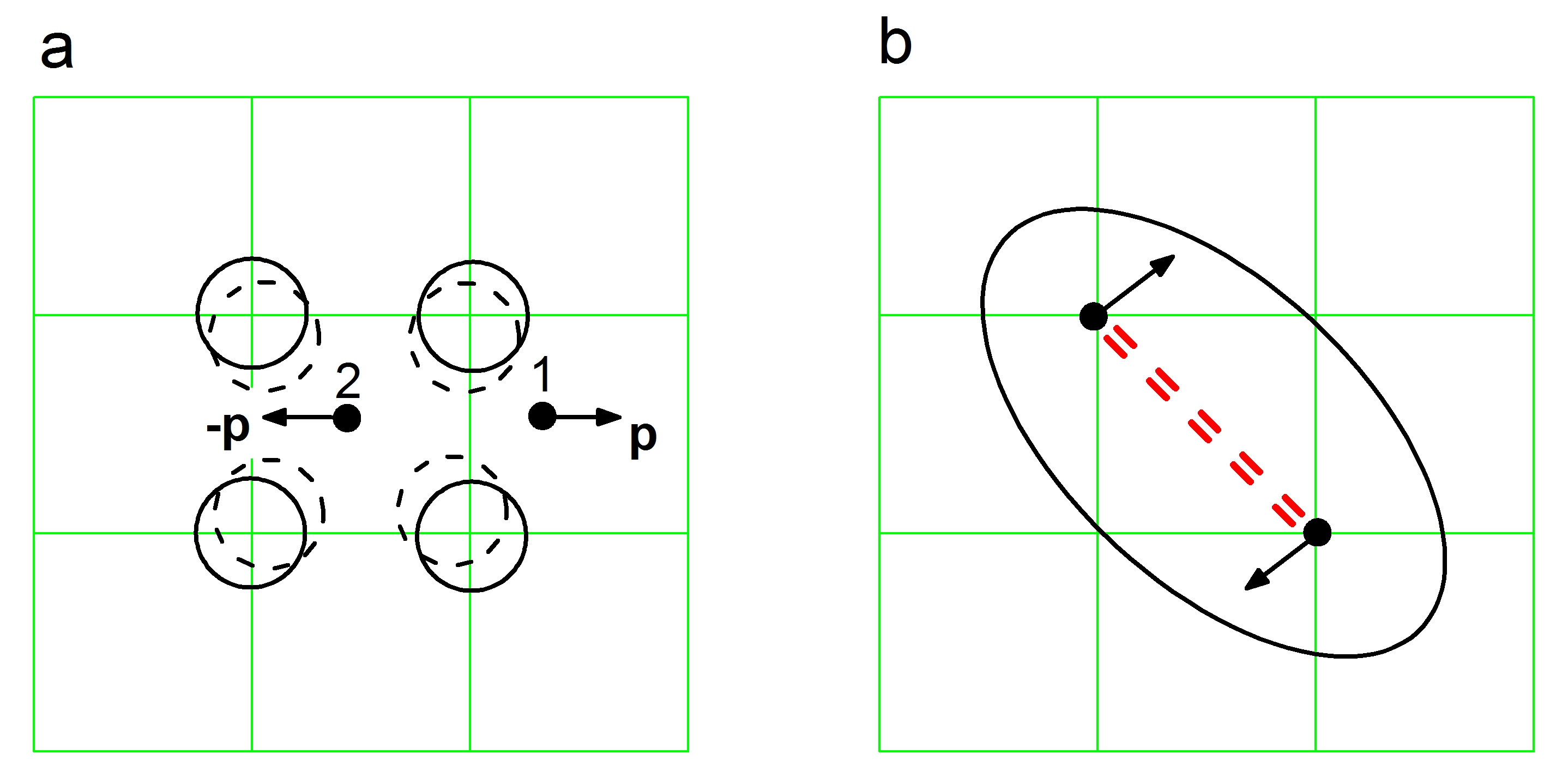}
\end{center}
\caption{
Illustration of superconducting pairing.
a. Conventional superconductivity. Ionic vibrations mediate attraction between two electrons
of opposite momenta. Electron 1 leaves a distortion of the ions (circles), which provides attraction
for electron 2 with opposite momentum. Because of a time delay, the Coulomb repulsion between
the electrons is avoided.
b. Unconventional superconductivity. In order to lower their energy, a pair of electrons avoids the
Coulomb repulsion by means of spatial anisotropy in their relative motion. In this process,
an attractive coupling (double-dashed line) is generated.
}\label{pairing}
\end{figure}

Superconductivity was discovered by H. Kamerlingh Onnes over a century ago \cite{Onnes}.
In a remarkable 1957 paper, the phenomenon was explained by J. Bardeen, L. Cooper
and J. R. Schrieffer(``BCS") \cite{BCS}.
Superconductivity
develops when electron pairs are condensed, that is, they move in unison and conduct
electricity  without experiencing any resistive loss of energy.
In the BCS theory, electrons form pairs through an attractive interaction mediated by phonons, the quantized ionic vibrations,
At the same time, Coulomb repulsion is avoided because a pair is formed between electrons with a time delay,
as shown in Fig.~1a.

In the FeSCs, superconductivity pairing does not arise from the conventional
electron-phonon coupling \cite{Boeri2008},
and is instead
a consequence of an unconventional
pairing mechanism generated by the electron-electron Coulomb interaction.
Qualitatively, electrons form pairs by an attractive force
produced in the process of avoiding
their Coulomb repulsion, as illustrated in Fig.~1b.

Two important characteristics of the FeSCs provide clues to the mechanism for their unconventional superconductivity:
In the FeSC phase diagram, superconductivity
emerges out of a ``bad-metal" normal state;
and the superconducting phase occurs near an
 AF order. This has led to extensive experimental and theoretical
 studies on the effect of  electron correlations and the
 nature of magnetism
 in the iron pnictides and chalcogenides. In the process,
 several significant properties have been discovered. One is
 electronic nematicity
 and its relationship to magnetism.
 Another concerns quantum
 criticality which arises
 at the border of the
 AF order. All these effects are also intimately connected to the amplitude and structure of the superconducting pairing.
	
The purpose of the present article is several fold. We first summarize the materials basis and electronic structures for the FeSCs. 
We then present a status report on key aspects of the microscopic physics.
At the present time, a variety of theoretical approaches are being taken to understand these systems.
Rather than describing any theoretical framework, we have organized this part of the article by considering the hierarchy of the significant energy scales in these materials:
Coulomb repulsion
with an order of magnitude of $1$ eV,
 followed by the energy scale for the antiferromagnetism,
of order  $0.1$ eV and reaching down to the superconductivity, whose pairing energy scale is 
on the order of $0.01$ eV.
Our conclusion includes a discussion of the prospects for further discoveries and understandings as well as the implications such studies have for the overall field of unconventional superconductivity.

\section{Materials basis and electronic structures}

FeSCs have  an extensive materials basis.
Fig.~2a illustrates the structure of several iron pnictides and chalcogenides.
A common feature in all these materials is the existence of either Fe-pnictogen
or Fe-chalcogen trilayers. In each FeAs/FeSe trilayer, the Fe ions form a square lattice,
and the As/Se atoms are located above or below the center of a square plaquette of the Fe ions.
 LaFeAsO belongs to the 1111
iron-pnictide family,
comprising FeAs trilayers that are separated by LaO layers.
The maximum $T_c=56$ K in the 1111 family is achieved in the SmFeAsO system
with
F-doping~\cite{ChenRen2008}.
BaFe$_2$As$_2$ belongs to the 122 iron-pnictide family \cite{Rotter.2008}.
Here, each unit cell contains two FeAs trilayers,
which are separated by a layer of Ba ions.
Superconductivity arises by chemical substitution into the undoped compound,
as illustrated in Fig.~2b for the case of Ni$_x$ substitution into BaFe$_2$As$_2$.
The maximum $T_c$ that has been reached in such 122 iron pnictides is 38 K, occuring in Ba$_{0.6}$K$_{0.4}$Fe$_2$As$_2$.

%%%% Figure 2%%%%
 \begin{figure}[!ht]
\begin{center}
\includegraphics[
angle=270,
%angle=0,
%width=0.85\textwidth]{Fig2RY922}\\[-38ex]
width=0.82\textwidth]{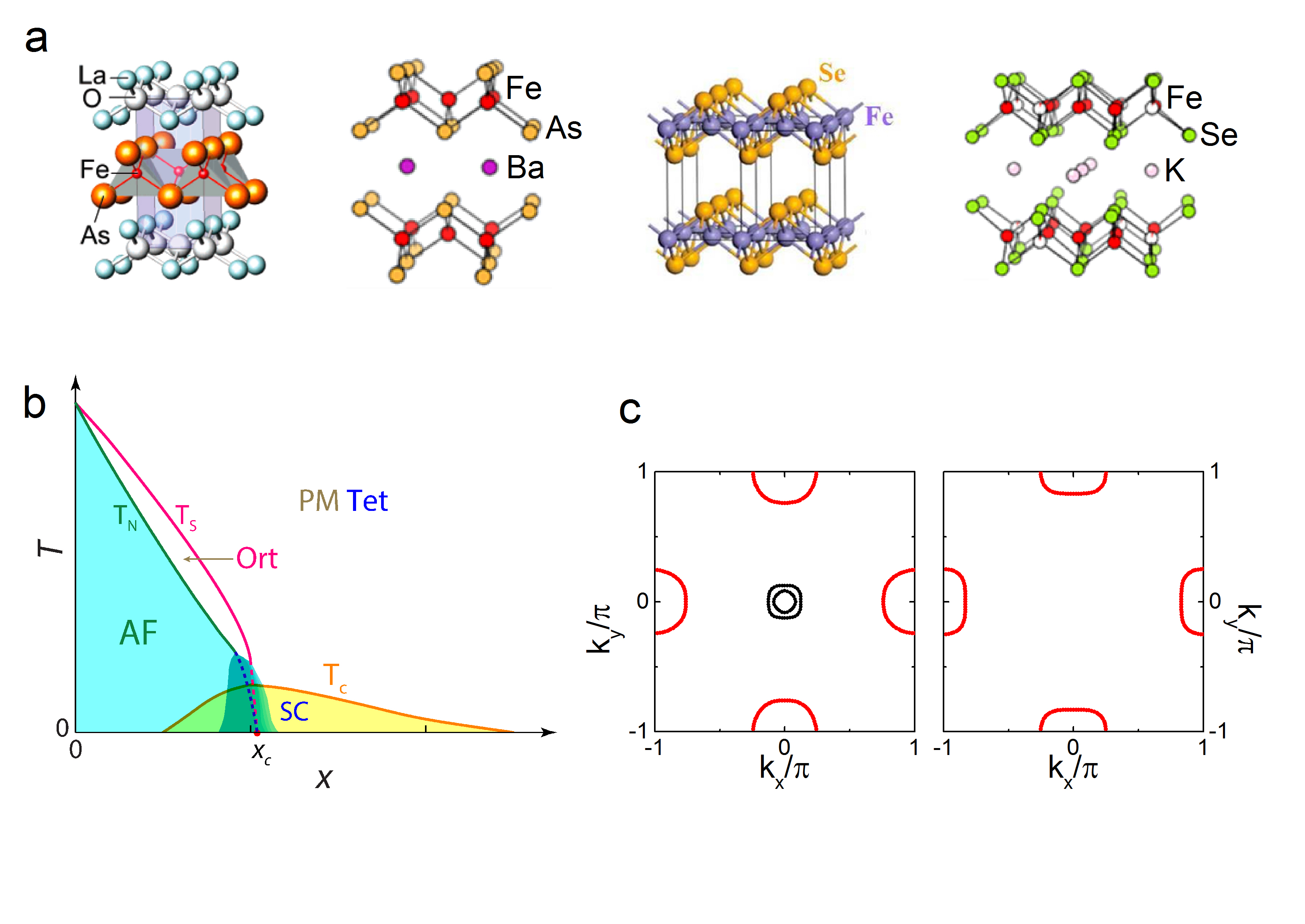}\\[-38ex]
%width=1.05\textwidth]{Fig2RY922}\\[-38ex]
\end{center}
\end{figure}
\newpage
\begin{figure}
\caption{
Materials characteristics of the iron-based superconductors.
a: Crystal structure of LaFeAsO (1111), BaFe$_2$As$_2$ (122),
FeSe (11) and K$_{x}$Fe$_{2-y}$Se$_2$.
Adapted from Refs.~\citenum{Hos2008},~\citenum{HosonoKuroki.2015}, and \citenum{HsuWu2008}.
b: Schematic phase diagram of BaFe$_{2-x}$Ni$_x$As$_2$ in the temperature ($T$)
 and chemical-substitution ($x$) plane.
 Adapted from Ref.~\citenum{XLu.2013}.
 c: Schematic Fermi surfaces
 of the iron pnictides (left panel)
 and those of several iron-chalcogenides
 (right panel, see main text).
 Adapted from Ref.\ \citenum{RYu.2013a}.
 }\label{fig:2}
\end{figure}
%%%%%%%%%%%%

Also shown in Fig~2a are two representative families of iron chalcogenides.
FeSe has the simplest structure in this category~\cite{HsuWu2008},
with each unit cell being an FeSe trilayer.
Each FeSe trilayer corresponds
to an Se-for-As replacement of an FeAs trilayer.
The potassium iron selenides \cite{JGuo.2010,LSun.2012}
K$_{x}$Fe$_{2-y}$Se$_2$, can be viewed
as derived from the FeSe system by inserting K ions between two FeSe trilayers.
In these bulk iron selenides,
the maximum $T_c$ that has been reached is similar to that of the 122 iron pnictides.
More recently,
``single-layer'' FeSe on SrTiO$_3$ substrate (``STO") has been
fabricated \cite{Xue.2012, Shen.2014}.
In this material,
there has been substantial evidence for superconductivity up to 65 K,
as indicated by the opening of a gap in the electron spectrum \cite{SLHe2013}
and the onset of  Meissner effect \cite{YWang2015},
as well as
a report of $T_c$ of
109 K,
as evidenced by
electrical transport measurements\cite{Jia.2014}.

Despite the similarities in their crystal structures,
an important feature for the FeSCs is a large variation of their electronic structures.
This is highlighted by the
Fermi surfaces. Fig.~2c illustrates the typical Fermi surfaces of the iron pnictides, comprising hole Fermi pockets
in the middle of the Brillouin zone and electron Fermi pockets at the boundaries of the Brillouin zone~\cite{Yi09}.
By contrast,
in several iron chalcogenides,
there are only electron Fermi pockets, as represented in the right-hand panel of Fig.~2c.
This describes single-layer FeSe on STO substrate. It also applies to K$_x$Fe$_{2-y}$Se$_2$,
which however
contains an additional electron pocket near the $\Gamma$ point in the Brillouin zone
(see later, Fig.~7b),
and
there is not any
hole Fermi pocket.
As yet another example,
in a class of extremely hole doped FeSCs
such as KFe$_2$As$_2$,
the zone boundary electron-like Fermi surfaces are absent
and have turned into
tiny hole Fermi pockets after a Lifshitz transition\cite{Sato09}.
We note that Fig.~2c illustrates the Fermi surfaces in purely two-dimensional (2D) Brillouin zone
of the one-Fe unit cell. In reality, they are only quasi-2D and ``warp" ({\it i.e.} disperse) along
 the third direction of the wavevector space, to varying
degrees in different iron-based materials.

\section{Electron correlations and bad-metal behaviour}

An important clue to the microscopic physics of the FeSCs comes from the early
observation that the metallic phases of the iron pnictides and chalcogenides
are all characterized by bad-metal properties, as we explain below.
Experimental and theoretical studies of
 such bad-metal behavior have provided considerable insights
into the degree of electron correlations in both these classes of materials.

\subsection{Iron pnictides and chalcogenides as bad metals}

Consider a representative iron arsenide, such as BaFe$_2$As$_2$.
It is metallic and develops AF order at a N\'eel temperature ($T_N$)
which is about 140 K.
\cite{ PCDai.2015}
As with the other iron-based compounds, above $T_N$ it is a paramagnetic metal with a rather large electrical resistivity. Several of these compounds are sufficiently clean to allow the observation of quantum oscillations at low temperature. However, even these have a very large magnitude of electrical resistivity at room temperature.
This signifies a bad metal by the Mott-Ioffe-Regel criterion \cite{Hussey2004},
namely the mean free path $\ell$ is very short,
on the order of inter-particle spacing:
its product with the Fermi wavevector, $k_F \ell$ is of order unity.
To estimate $k_F \ell$,
one recognizes that there are $p=4$ or $5$ bands crossing the Fermi level (Fig.~2c).
Based on
the single-crystal in-plane resistivity at room temperature
of about 400 $\mu$Ohm-cm, the estimated $k_F\ell \approx 5/p \approx 1$
for each Fermi pocket.
This should be contrasted to a good
metal, such as Cr, for which the resistivity in the paramagnetic state just above the
room temperature is about 40 times smaller\cite{Fawcett-RMP}.
Because electron-phonon scattering has a much smaller contribution to the resistivity,
such a small value of $k_F \ell$
in the iron pnictides
signifies strong electron-electron
interactions\cite{SiAbrahams.2008,Si-NJP2009}.

A related signature of the strong electron correlations is a considerable
 reduction
in the weight of the Drude peak of the optical conductivity.
As illustrated in Fig.~3a,  Qazilbash  {\it et al}\cite{Basov.2009}
observed
that the spectral weight in the Drude peak
is about $30$\%
of the value expected from non-interacting electrons.
This reduction is quite sizable, on the same order as that seen
in other bad metals such as the normal state of the doped cuprate superconductors
and V$_2$O$_3$, where correlation effects are of primary importance.
A corollary is that about $70$\% of the spectral weight resides in an incoherent part,
associated with electronic states
away from the Fermi energy.
Similar behavior has also been seen in
various
other
iron pnictides \cite{Nakajima.2014}.

%%%% Figure 3%%%%
 \begin{figure}[!b]
 %[!ht]
\begin{center}
\includegraphics[
angle=0,
%width=0.95\textwidth]{Fig3RY1210}\\[-38ex] %Fig3RY1210
width=0.86\textwidth]{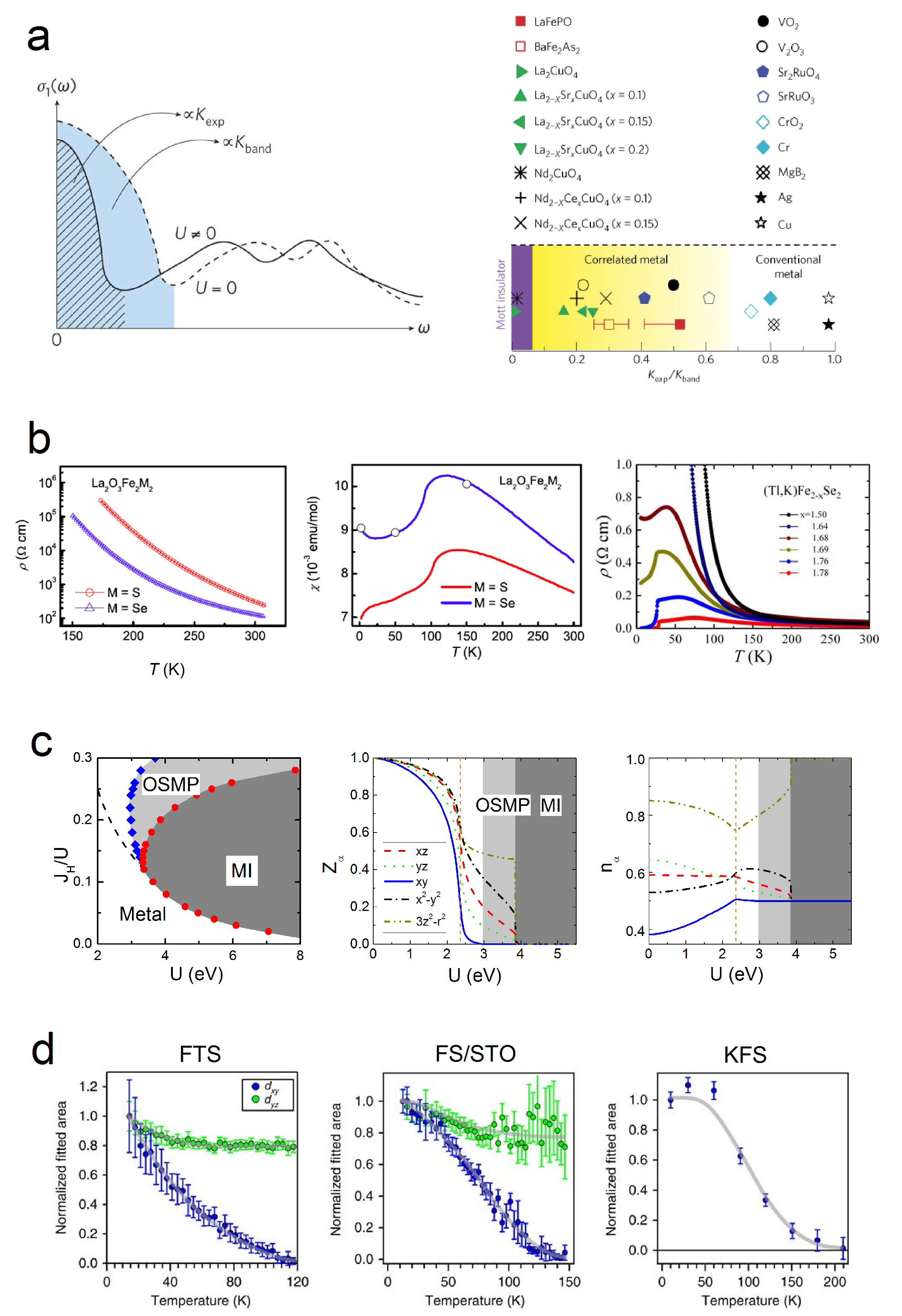}\\[-38ex] %Fig3RY1210
\end{center}
\end{figure}
\newpage
\begin{figure}[t!]
\caption{\
Bad-metal behaviour and electron correlations. a: Left panel: Gauging
electron correlations
by the ratio of $K_{exp}$, the
measured
Drude weight
of the optical conductivity $\sigma_1(\omega)$ vs. frequency $\omega$,
to $K_{band}$, its non-interacting
counterpart. Adapted from Ref. \citenum{Si-NatPhys.2009}.
Right panel: $K_{exp}/K_{band}$ for various systems, including the iron arsenides.
 From Ref. \citenum{Basov.2009}.
b: Insulating iron chalcogenides.
Left and middle panels: resistivity ($\rho$) and magnetic susceptibility ($\chi$)
{\it vs.} temperature $T$ for the oxychalcogenides.
From Ref.~\citenum{Zhu2010}.
Right panel: $\rho$ {\it vs.} $T$ for the alkaline iron selenides.
From Ref.~\citenum{Fang2011}.  \;c:
Left panel: Ground-state phase diagram for the alkaline
iron selenides at
$N=6$
(see the main text).
The dark and light grey regions correspond
to the Mott insulator (MI) and OSMP, respectively.
The black dashed line describes a crossover into a strongly-correlated metal regime.
Right panel: In the OSMP, the quasiparticle weight $Z$ vanishes
for the $3d$ $xy$ orbital but is
nonzero for the others; several orbitals are close to half-filling.
From Ref.~\citenum{Yu2013}.
d: 
Quasiparticle weight near the Fermi level determined from ARPES data for bands with distinct
%$3d$ 
$xy$ and $yz$ characters,
providing evidence
for the OSMP in
FeTe$_{0.56}$Se$_{0.44}$ (FTS, left panel),
monolayer FeSe film on SrTiO$_3$ (FS/STO, middle panel),
and K$_{0.76}$Fe$_{1.72}$Se$_2$ (KFS, right panel).
From Ref.~\citenum{MYi.2015}.
}\label{fig:3}
\end{figure}

The reduction of the Drude weight
is accompanied by the
mass renormalization observed in the angle-resolved photoemission spectroscopy
(ARPES). In the iron arsenides,
$m^*/m$
in the paramagnetic phase
is
about $3$--$4$.\cite{Yi09,Yi_NaFeAs12}
In the iron chalcogenides, in which the room-temperature resistivity also reaches the
Mott-Ioffe-Regel limit, the effective mass
enhancement can be as large as $20$ for some of the involved bands
\cite{Tamai10,MYi.2013,MYi.2015}.

These bad-metal properties suggest that the electron-electron correlations are sufficiently
strong to place the metallic iron pnictides and chalcogenides
in proximity of a Mott localization.
Can we tune up the strength of the electron correlations
and thereby push these and related materials into the Mott insulating state?

\subsection{Mott insulators in the iron chalcogenides}
A characteristic measure of the strength of the electron correlations
is the ratio of the local electron-electron interaction to the electron bandwidth, or the kinetic energy.
Therefore, reducing the kinetic energy would effectively enhance the correlation effects.
This led to the consideration of the iron oxychalcogenides
La$_2$O$_2$Fe$_2$O(Se,S)$_2$.
Like the FeAs trilayer of the iron pnictides,
each Fe$_2$O(Se,S)$_2$ layer
contains a square lattice of Fe-ions, which also have a nominal valence of $+2$.
Howver, the Fe-Fe distance has been increased, thereby leading to a reduced
kinetic energy.
Calculations of their electronic structure demonstrated that their
Fe $3d$-electron bandwidth is
about 75\%
of that
in the usual cases such as
LaFeAsO and BaFe$_2$As$_2$;
the consequently enhanced electron correlation effects raised the possibility for a Mott insulating state
in these materials \cite{Zhu2010}. Experimentally, their insulating behavior
is well-established
by the temperature dependence
of electrical resistivity, which shows an
activated behavior
with moderate charge gaps ($0.19$ eV and $0.24$ eV, respectively for the
Se and S cases) as shown in Fig.~3b, left panel.
AF orderings occur around $T_N$ of about
$93$ K and $105$ K for the Se and S cases, respectively, as indicated by the
temperature dependence
of the magnetic susceptibility (Fig.~3b, middle panel) and neutron-scattering
studies\cite{neutrons}. The insulating behavior persists above $T_N$, a characteristic
signature of a Mott insulator\cite{Imada-RMP}.
Finally, the incoherent electronic excitations of such a correlated insulating state have
recently been observed by X-ray spectroscopy \cite{Freelon}.

Insulating behavior has also been observed in several other iron chalcogenides.
One such case is the alkaline iron selenides \cite{Fang2011}, as shown in Fig.~3b,
right panel.
This corresponds to a ``245" composition
A$_{0.8}$Fe$_{1.6}$Se$_2$ (A=K,Tl/K,Rb),
in which one out of five Fe-atoms is absent
in each FeSe trilayer \cite{Bao.2011}.
It has been suggested that the ordered vacancies reduce
the kinetic energy and, in a way similar to the oxychalcogenide case described above,
gives
rise to a Mott insulating state \cite{Yu2011,Zhou2011}.
The alkaline iron selenides  have a ``234" phase as well, AFe$_{1.5}$Se$_2$, for
which too the Mott insulator nature
has
been suggested
based on
measurements with
angle-resolved photoemission spectroscopy
(ARPES) \cite{Birgeneau2015}.

\subsection{Orbital-selective Mott physics}
\label{subsec:osmp}

To study the effect of electron correlations, it is important to recognize the multi-orbital
nature of the electronic states.
Conversely,
the multi-orbital physics can be
utilized to
extract clues about the degree and nature of the electron correlations.
BaFe$_2$As$_2$ is an example of a parent compound, ({\it i.e.} undoped, stochiometric)
in which the Fe valence is $+2$. In these compounds,
there are $N=6$ electrons on average that occupy the
five $3d$ orbitals of each Fe ion.
The theoretical description then is given by multi-orbital Hubbard models, with the
minimal interactions being a Hubbard interaction $U$ and a
Hund's coupling $J_H$.
Such models typically
include at least the $3d$ $xy$ and $xz/yz$ orbitals,
and
can involve all the five $3d$ orbitals.
The interplay of the kinetic energy,
the short-range repulsion $U$ and the Hund's coupling $J_H$
may lead
to different behaviors
for the various orbitals. It may even be possible that a subset
of the orbitals undergoes
 Mott localization, thus orbital-selective Mott physics.

ARPES measurements have
provided  evidence \cite{MYi.2013,MYi.2015} for an orbital-selective Mott phase (OSMP).
This is illustrated in Fig.~3d, which shows that, for each of the three iron chalcogenides,
as temperature goes
above about $100$ K,
the spectral weight for the
$3d$ $xy$ orbital has vanished, while that for the
$3d$ $xz/yz$ orbitals has not changed much.
The experimental results suggest that this regime can be described by an OSMP, in which the $3d$ $xy$ electrons
are localized while those associated with the other $3d$ orbitals remain itinerant.
 Additional evidence for the OSMP has come from
THz spectroscopy \cite{Wang2014}, Hall measurements \cite{Ding2014},
pump-probe spectroscopy \cite{Li2014} and high-pressure
transport measurements \cite{Gao2014}.

The OSMP was anticipated theoretically. Fig.~3c shows the zero-temperature phase
diagram calculated for the potassium iron chalcogenides at the $3d$-electron filling
$N=6$, in the paramagnetic case with tetragonal lattice symmetry
(instead of the AF-ordered case, to highlight the localization effects associated
with the Coulomb repulsive interactions).
When the combined $U$ and $J_H$ interactions are sufficiently large,
a Mott insulator phase arises.
An OSMP  occurs
between this phase and the one in which all the
$3d$ orbitals are itinerant. \cite{Yu2013,Yu-COSSMS2013}.
When the electron filling deviates from the commensurate value $N=6$,
the Mott insulator phase is suppressed but the OSMP survives a finite range
of electron filling.
Similar results arise for a model without the 
ordered Fe-vacancies.\cite{Yu2013}
When the system at zero temperature is purely itinerant but close to the
boundary of the OSMP, increasing temperature induces a crossover
 to an OSMP,
 which is consistent with experimental observations (Fig.~3d).

Several factors are responsible for the OSMP. First, in the noninteracting limit,
the bandwidth of the $xy$ orbital is smaller
than those of other orbitals. Second, the Hund's coupling suppresses
inter-orbital correlations;
this effectively decouples the $3d$ $xy$ orbital from others and
keeps the $xy$,$xz$ and $yz$ orbitals effectively
at half-filling.
(Such an inter-orbital decoupling is also essential for the stability of the
OSMP phase, given that the orbitals are mixed in the non-interacting limit.)
Third, the degeneracy of the $xz$ and $yz$
orbitals makes
 the threshold
interaction needed for their localization to be larger than its counterpart for the
non-degenerate $xy$ orbital\cite{Anisimov2002}.
As a combined effect of
these factors, the $3d$ $xy$ orbital has a lower interaction
threshold for the Mott transition,
and the OSMP  ensues.
We note that
the OSMP is an extreme limit of the effects of orbital-selective correlations.
More generally, the $3d$ $xy$ orbital is not Mott localized,
but  is close to a Mott localization, and the system may still exhibit strong orbital
dependent effects\cite{Yu.2012,deMedici2014,Bascones}.
We also note that OSMP is pertinent to other classes of strongly correlated
electron systems \cite{Anisimov2002,Neupane2009}.

The experimental observations and general considerations
summarized in this section
suggest the importance of electron-electron correlations in determining the properties
of the iron pnictides and chalcogenides. They have provided the basis
for various approaches to these systems
\cite{SiAbrahams.2008,RYu.2013a,Yu.2012,deMedici2014,Bascones,Yin2011,Fang08,Xu08,Laad.2009,KSeo,Moreo.2009,WQChen.2009,DHLee.2013,Berg.2009,Lv,Yu2015,FWang2015}
which are ``strong coupling", in the sense that the effect of electron-electron interactions
is treated non-perturbatively.

\section{Magnetism and electronic nematicity}

Because an AF order typically exists near superconductivity
in the phase diagram of the
FeSCs, magnetism has received considerable attention in the field
since the very beginning.\cite{Cruz2008}
There has also been increasing recognition for the role that nematic order, the breaking
of orientational symmetry,
plays in uncovering the microscopic physics.

\subsection{Magnetism in the iron pnictides}

As already illustrated in Fig.~2b, the parent iron pnictides are antiferromagnetically ordered.
The ordering wavevector is $(\pi,0)$ (the notation is that of the Fe square lattice),
as shown in Fig.~4a.
This AF
order
 is the background for quantum fluctuations of the spins
 below the N\'eel transition  temperature $T_N$.
These have been probed by inelastic neutron scattering
\cite{PCDai.2015},
which measures the
frequency and wavevector dependences of the spin structure factor,
$S({\bf q},\omega)$, or  of the dynamical spin susceptibility, $\chi^{\prime\prime}({\bf q},\omega)$.

A striking feature
 is that the spin fluctuations remain very strong
over a wide temperature  range above $T_N$.~\cite{Diallo10,Harriger11,Ewings11}
In Fig.~4a,  the dynamical spin structure factor
 is shown in terms of equal-intensity contours
in the wavevector space. At relatively low energies, these are ellipses
near $(\pm \pi,0)$ and $(0, \pm \pi)$. At high energies, they have the form of
spin-wave-like excitations all the way to the boundaries of the
AF Brillouin zone~\cite{Harriger11}. The peak intensity traces out a
spin-wave-like energy dispersion, which is also shown in Fig.~4a.

The electron-correlation effects implied by the bad-metal behavior
discussed above have
inspired the study of the magnetism with the local moments as the starting point.
Electron correlations turn the majority of the single-electron excitations incoherent
and distribute them away from the Fermi energy. An expansion has been
constructed in terms of the fraction, denoted by $w$,
of the single-electron spectral weight that lies in the low-energy
coherent part of the spectrum.~\cite{SiAbrahams.2008,Si-NJP2009}.
To the zeroth order in $w$,
the incoherent electronic excitations
(when their charge degrees of freedom are ``integrated out")
give rise to
localized magnetic moments associated with the Fe ions.
In the case of the iron arsenides, the
 $p$-orbitals of the
 As ions
 mediate the
exchange interactions among the local moments.
 This leads to geometrical frustration in the magnetism:
With each As ion sitting at an equal distance from the Fe ions
of a square plaquette (Fig.~2a), the exchange interactions contain
 $J_1$ and $J_2$,
between the nearest-neighbor and next-nearest-neighbor sites, respectively,
on the
Fe-square lattice ~\cite{SiAbrahams.2008,Yildirim.2008,Ma2008}.
General considerations suggest that $J_2
>  J_1/2$.
In this parameter regime,
the ground state of the $J_1$-$J_2$ Heisenberg model on the square lattice
has the collinear $(\pi,0)$ order \cite{Chandra90}, as is observed experimentally.

Because the charge gap in the incoherent excitations
 is relatively small, and also due to the Hund's coupling in
the multi-orbital setting \cite{Fazekas} of the Fe-based compounds,
multiple spin couplings may be of importance. In particular, the four-spin biquadratic
coupling $K$ of the form $K({\bf S}_i\cdot{\bf S}_j)^2$
has been shown\cite{JJK,Wysochi.2011} to be significant.
Moreover, at higher order in $w$, there are itinerant
coherent electrons which are coupled to the local moments and introduce damping
to the spin excitations. The calculated spin structure factor~\cite{JJK}
 for the
$J_1$-$J_2$-$K$ model model in the presence of damping are displayed in Fig.~4b.
The theoretical
results provide a consistent understanding of the experimental data.

As the majority of the electron spectral weight is in the incoherent sector, the above discussion
implies a large spin spectral weight. It has become possible to test this experimentally.
In Fig.~4c, $\chi^{''}$, the momentum-integrated dynamical spin susceptibility \cite{Liu.2012}
as a function of energy
$E$ is shown. A further integration over energy, corresponding to
$\int d E \int d {\bf q}\, S({\bf q},E)$,
 yields a total spectral weight of about  $3\mu_B^2$ per Fe.
 This corresponds to the spin spectral weight
 of a full spin-$1/2$ moment, {\it i.e.} one $\mu_B$ per Fe, as anticipated.

The spin spectral weight and its distribution in energy provide another means to characterize
the strength of electron correlations. For weak correlations, the spin degrees of freedom should
be described in terms of triplet excitations of electrons and holes near the Fermi energy.
In such a weak-coupling description~\cite{Dong08,Knolle11},
the enhancement of the spin excitations  near $(\pm \pi, 0)$ and $(0,\pm \pi)$
would arise from a Fermi-surface ``nesting" effect, {\it i.e.}
the enhanced phase space for connecting the electron and hole
Fermi pockets near these wave vectors.
However, the small size of the electron and hole
Fermi pockets would limit their contributions to the spin spectral weight to a considerably
smaller value than what is observed via the integrated dynamical susceptibility.

 %%% Figure 4%%%%
\begin{figure}[!b]
\begin{center}
\includegraphics[
angle=0,
width=0.83\textwidth]{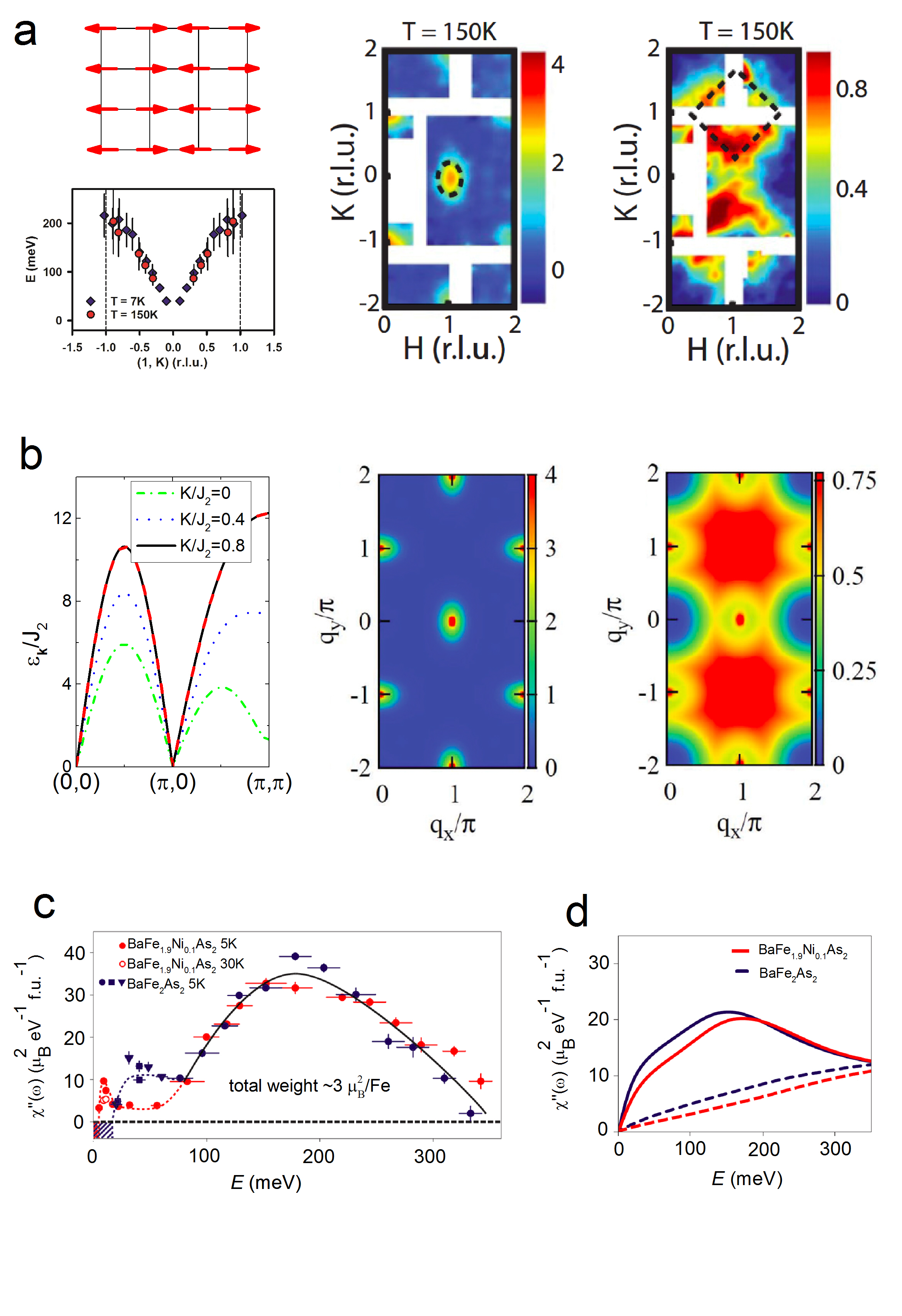}\\[-38ex]
\end{center}
\end{figure}
\newpage
\begin{figure}[!t]
\caption{Magnetism in the iron pnictides. a. Left top panel: the magnetic ordering pattern of the
parent iron pnictides such as BaFe$_2$As$_2$. Left bottom, middle and right panels:
the energy dispersion and spin structure factor measured in BaFe$_2$As$_2$
above the N\'eel temperature. The wavevector space is denoted by
(H,K) (r.l.u.)=(H$\pi$,K$\pi$) in our notation.
Adapted from Ref.~\citenum{Harriger11}. b. Theoretically calculated
energy dispersion and spin structure factor within a $J_1$-$J_2$-$K$ model. Adapted from
Ref.~\citenum{JJK}. Also shown are the imaginary part of the dynamical local spin
susceptibility
measured in pure and Ni-doped BaFe$_2$As$_2$ (c)
and calculated from the dynamical
mean field theory with strong correlations (solid lines) and from
the weak-coupling RPA method (dashed lines) (d) adapted from Ref.~\citenum{Liu.2012}.
}
\label{fig:4}
\end{figure}
%%%%%%%%%%%%

A quantitative analysis is shown in Fig.~4d. Calculations based on the dynamical
mean-field theory (solid lines),
which incorporate the contributions from the incoherent part of the electron spectral weight,
capture the right size of the spin spectral weight within the energy range
of experimental measurements \cite{Liu.2012}. A weak-coupling calculation,
based on the random-phase approximation (RPA) (dashed lines)
considerably underestimates the spin spectral weight.

\subsection{Magnetism in the iron chalcogenides}

An AF order also appears in a variety of iron chalcogenides. In FeTe,
it has the pattern illustrated in Fig.~5a,
corresponding to an ordering wavevector
$(\frac{\pi}{2},\frac{\pi}{2})$. This ordering pattern
can be understood in terms of local moments coupled through
multi-neighbor $J_1$-$J_2$-$J_3$ exchange interactions,
in the presence of the biquadratic $K$ couplings~\cite{Ma2009,Yu2015}.
The spin spectral weight,
$\int d E \int d {\bf q} \,S({\bf q},E)$,
 is now even larger,
given that the ordered moment
is already $2.5~\mu_B$ per Fe.~\cite{Wen2015}

These observations further elucidate the microscopic physics.
 The Fermi surface
in the paramagnetic state of the 11 ({\it e.g.} FeTe) iron chalcogenides
is similar to that of the iron pnictides,
as illustrated in Fig.~2c, left panel.
For such a Fermi surface, $(\frac{\pi}{2},\frac{\pi}{2})$ is not a nesting wavevector.
Therefore, the magnetic ordering pattern in FeTe cannot be understood within
an itinerant
nesting description.
This suggests that the above considerations on the spin spectral weight of the iron pnictides
continue to apply in the FeTe case.

%%% Figure 5%%%%
%\begin{figure}[!ht]
\begin{figure}[!b]
\begin{center}
\includegraphics[
angle=270,
%angle=0,
%width=0.8\textwidth]{Fig5RY1203}\\[-38ex]
width=0.70\textwidth]{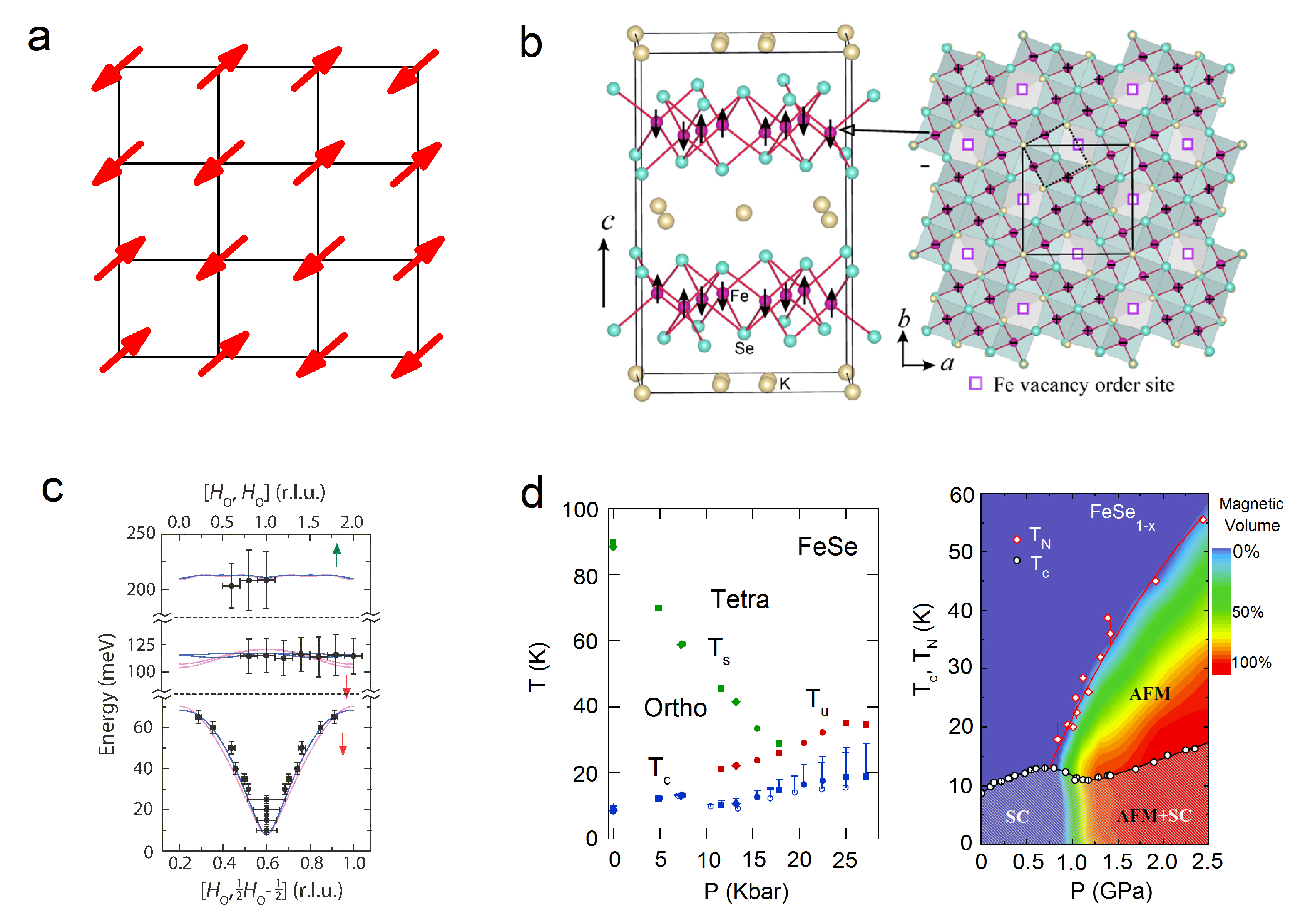}\\[-38ex]
\end{center}
\end{figure}
\newpage
\begin{figure}
\caption{Magnetism in the iron chalcogenides. a. The magnetic ordering pattern
of FeTe. b. The block-spin AF order in the 245 K$_2$Fe$_4$Se$_5$
 (from Ref.~\citenum{Bao.2011})  c. The corresponding spin-wave dispersion (from
 Ref.~\citenum{MWang2011})  d. Phase diagram of bulk FeSe, showing
 the structural transition and pressure-induced superconductivity (left panel, from Ref.~\citenum{Terashima15})
 and the pressure-induced magnetic order (right panel, from Ref.~\citenum{Bendele12}).
}
\label{fig:5}
\end{figure}
%%%%%%%%%%%%

The 245 ({\it e.g.} K$_2$Fe$_4$Se$_5$) alkaline iron selenides have a block spin AF order,
as illustrated in Fig.~5c. Again, the ordered moment is large, about
$3.3~\mu_B$ per Fe.
This magnetic order can be understood in terms of an extended
$J_1$-$J_2$ model on a $\frac{1}{5}$-depleted square lattice with
$\sqrt{5} \times \sqrt{5}$ vacancy order~\cite{Yu-Goswami-Si2011,Cao-Dai2011}.
Fig.~5c illustrates the four branches of the observed
spin-wave spectra and a theoretical fit based on
the local-moment model. The success of this local-moment description of the
magnetic dynamics, together with the large ordered moment, provide evidence
for the Mott-insulating nature of the 245 alkaline iron selenides.
It can also be seen from the momentum distribution of the spin spectral weight
~\cite{MWang2011,Chi13}
that the damping rate here is smaller than that inferred in the iron pnictides.
The overall behavior is similar for the 234 alkaline iron selenides
({\it e.g.} Rb$_{0.8}$Fe$_{1.6}$Se$_2$),
but the
different vacancy pattern
 promotes a different AF order~\cite{MWang2013}.

 \subsection{Electronic nematicity and its relation with magnetism}

The iron pnictides typically have parent compounds whose ground states have both a collinear $(\pi,0)$ AF order
as well as a structural distortion \cite{PCDai.2015}, as illustrated in Fig.~2b. 
Resistivity anisotropy measurements
under uniaxial stress \cite{IFisher.2012,HHKuo2015} above the structural transition in 122 compounds have revealed
a very large electronic nematic response and an examination of the strain dependence of the resistivity anisotropy
(proportional to the electronic nematicity) 
gave compelling evidence that the structural transition
is electronically driven. Other probes
of the nematic correlations \cite{bohmer_shear_2014,thorsmolle2014,kretzschmar2015},
while not as direct, have given results consistent with that that conclusion.

One explanation for the
electronic-nematic transition
is
 a magnetic Ising-nematic
order.
It was
recognized from the beginning that models with quasi-local moments
and their frustrated Heisenberg $J_1$-$J_2$ interactions
\cite{SiAbrahams.2008} feature such an Ising-nematic transition
\cite{Fang08,Xu08,JDai.2009,Chandra90}, and similar conclusions have
subsequently been reached in models
based on Fermi-surface instabilities
\cite{Fernandes14}.
While the magnetic mechanism for the nematicity has had considerable
success,
it is yet to be unequivocally established.
An alternative proposal attributes the origin of the nematic order to the orbital
degrees of freedom.\cite{Lv,Devereaux10, Lee2009, Kruger2009}.
Indeed, ARPES measurements in uniaxially pressurized  BaFe$_{2-x}$T$_x$As$_2$
 show a splitting between the two orthogonal bands, which are otherwise degenerate, that have
 dominant $d_{xz}$ and $d_{yz}$ characters \cite{MYi.2011}.
Adding to the complication is that
symmetry allows a bilinear coupling between the magnetic Ising-nematic
and orbital order parameters, and care must be exercised to sharpen the distinction
between the scenarios.

Along this direction, two recent developments are worth emphasizing.
Neutron scattering experiments have measured
the spin-excitation
anisotropies~\cite{Lu-Science2014,Song2015}.
Consider the collinear AF order for the iron pnictides.
 Applying a uniaxial strain along one axis of the orthorhombic lattice
 yields a spin excitation anisotropy, as measured by
$S_{\rm diff}(E) \equiv S[(\pi,0),E)-S[(0,\pi),E]$ in comparison
with
$S_{\rm sum}(E) \equiv S[(\pi,0),E)+S[(0,\pi),E]$.
Above the structural transition ($T>T_s$), experiments reveal a strong spin excitation anisotropy
in the optimally doped
regime, where the ARPES determined
 splitting between the $d_{xz}$ and $d_{yz}$ bands is already considerably
 diminished.
 Moreover quantitatively, while the ARPES-measured orbital splitting energy has gone from
 about 60 meV in the undoped BaFe$_2$As$_2$ to about 20 meV near the optimal
 electron doping, the energy scale for the spin excitation anisotropy has remained on the order
 of about 60 meV. These results suggest the dominant role that magnetism plays
 in leading to the nematic correlations~\cite{Song2015}.

 Another recent development concerns bulk FeSe.
As illustrated in Fig.~5d, this compound displays a tetragonal-to-orthorhombic
structural transition with $T_s \approx 90$ K, but no N\'{e}el transition
has been detected \cite{McQueen09,Medvedev09,Bohmer14,Baek14}.
It has been suggested that these results
imply a failure of the magnetism-based origin for the
structural phase transition\cite{Bohmer14,Baek14}.
There is, however, a natural way to understand this behavior within the magnetic
picture.
Several groups have studied the frustrated magnetism
associated with the spin-exchange interactions among the local
moments \cite{Yu2015,FWang2015,Glasbrenner2015}.
Based on the theoretical phase diagram associated with
the frustrated bilinear-biquadratic exchange interactions,
Ref.~\citenum{Yu2015}  proposed  that
the structural transition
in FeSe originates from an Ising-nematic order of an antiferro-quadrupolar
phase. This kind of order has the spins preferring an axis
without developing an orientation along it, thereby breaking the spin
rotational invariance while preserving the time-reversal symmetry.
While no static AF order occurs,
 the collective modes of this quadrupolar state yield
$(\pi,0)$ magnetic fluctuations,
which have since been observed by inelastic neutron scattering
measurements.\cite{Rahn15,WangZhao15}

\section{Quantum criticality}
\label{sec:qcp}
In many
correlated-electron
materials,
several ground states can be in competition and it is often possible  to tune from
one to another through a ``quantum critical point" (QCP)
by adjusting some control parameter, for example pressure, magnetic field,
chemical composition. An essential characteristic is that
dynamical fluctuations
of an order parameter
play a key role in determining the behavior in the neighborhood of the
QCP. Although quantum phase transitions between distinct ground states occur
at absolute zero, their effects may be observed over a range of  non-zero temperature.
The physics of quantum criticality in the iron pnictides and chalcogenides
has been reviewed in Ref.\ \citenum{AS}.

As the iron-based superconductors exhibit a number of different phases:
paramagnetic metal, superconductor, antiferromagnet in their phase diagrams,
it is reasonable to expect that it is possible to access one or more
QCPs
by identification of appropriate control parameters. Fig.~6a shows an example,
in which the QCP  separates an
AF
metal from a paramagnetic one,
and is  accessed by a tuning parameter, which, as we discuss below for
a particular compound, can be precisely the $w$ of the
measure of electron correlations that was introduced in Sec.\ IIIA.

The considerations of Section II on electronic correlation and proximity
to a Mott transition imply that by tuning the ratio  of kinetic energy
to Coulomb repulsion from small to large, one could pass
 from an  AF state to a paramagnetic one via a magnetic
 QCP. Fig.~6a  is adapted from Ref.\ \citenum{JDai.2009}, where it was first proposed
 that isoelectronic substitution of P for As in, for example
 LaFeAsO
 or BaFe$_2$As$_2$,
 would increase that ratio, here
 measured by $w$. This is because
  P is substantially smaller than As, which leads to a reduction
  in the unit cell volume (as well as the pnictogen height), thus an increase in kinetic energy.

  Apart from its intrinsic interest as an exploratory tool for the physics of quantum criticality,
the isoelectronic tuning of $w$ provides a  window on the effects of electron correlation
that is complementary to the studies described in previous sections,
which placed the iron chalcogenides on the side of stronger correlation
as compared to the pnictides.
Here we move to weaker correlation in the arsenides by P for As substitution.

The proposed phase diagram of Fig.~6a shows a quantum critical region,
in which thermodynamic and transport properties are expected to have power-law dependences
on temperature and tuning parameter. The
ordered phase is an antiferromagnet,
which disappears into a paramagnetic Fermi-liquid phase as one passes through the QCP.
Crossovers out of the region of quantum criticality are denoted by the dashed red lines.
The blue lines represent the thermally-driven AF
 transition (solid) and
 an Ising-nematic transition manifested through a structural transition (dashed).
The thermal transitions are shown to be two split second-order transitions, but may also be
concurrent first-order transitions.

%%% Figure 6%%%%
\begin{figure}[hb]
\begin{center}
\includegraphics[
%angle=270,
angle=0,
width=0.82\textwidth]{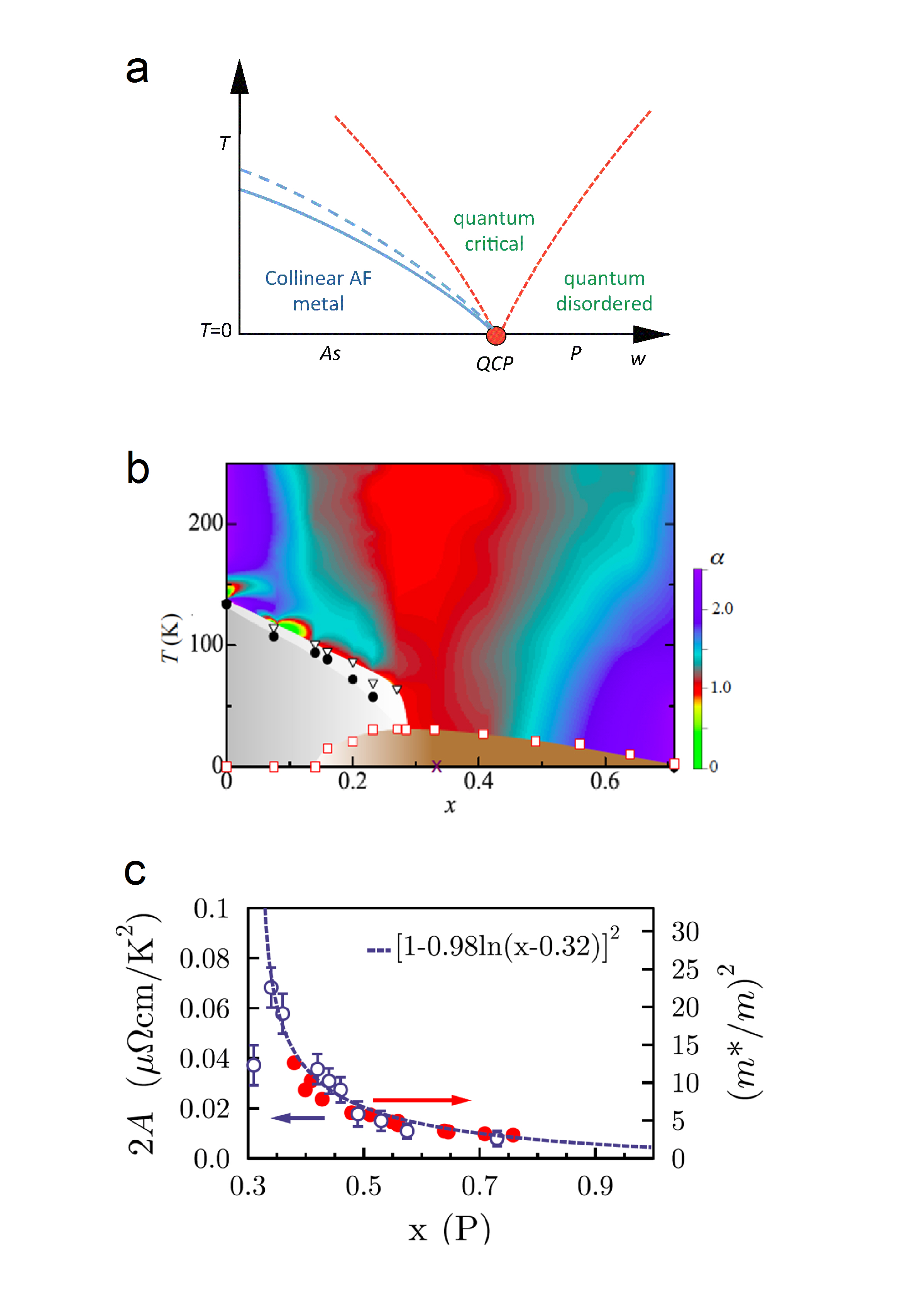}\\[-38ex]
\end{center}
\end{figure}
\newpage
\begin{figure}
\caption{
Quantum criticality in the iron pnictides. a. Quantum criticality in the theoretically proposed phase diagram.
The red dot marks a quantum critical point.
Adapted from Ref.~\citenum{JDai.2009}. b. Phase diagram of BaFe$_2$(As$_{1-x}$P$_x$)$_2$.
The color shading describes the variation of the resistivity exponent, $\alpha$,
in its temperature dependence,
represented by $\rho = \rho_0 + A T^{\alpha}$.
The cross marks the QCP at $x_c = 0.33$.
From Ref.~\citenum{YMat};
courtesy of Y. Matsuda.
c. Divergence of $m^*$, the effective mass, and $A$, the $T^2$-coefficient
of the electrical resistivity, on approach
of the QCP.
Adapted from Ref.~\citenum{Analytis.2014}, containing data of Ref.~\citenum{Walm};
courtesy of J. Analytis.
}
\label{fig:6}
\end{figure}

Experimental verification of quantum criticality induced by isoelectronic
substitution in order to reduce the correlation effects came
soon after the original proposal. This is illustrated in Figs.~6b and 6c.
In the former,  adapted from Ref.~\citenum{YMat}, the color shading represents
the resistivity  temperature exponent  $\alpha$ in $\rho=\rho_0 + AT^{\alpha}$
in BaFe$_2$(As$_{1-x}$P$_x$)$_2$ (122 compound),
in which the QCP occurs at a critical value of $x=x_c\approx 0.33$.
The red area indicates the quantum critical regime in which the resistivity
has a linear $T$-dependence as is characteristic \cite{LRVW}
for a two-dimensional
AF QCP.
Furthermore, in the approach to the QCP, the expected singular behavior
of thermodynamic quantities is observed as is seen in Fig.~6c,
adapted from Ref.\citenum{Analytis.2014} which shows
the expected \cite{LRVW} logarithmic increase of the electron effective mass
as $x\rightarrow x_c$. The left-hand scale shows the measured
``$A$ coefficient" of the $T^2$ resistivity in the quantum-disordered region,
where it should vary as $(m^*)^2$ \,\cite{KW}.
This figure demonstrates a consistent quantum critical behavior
between transport and thermodynamic measurements {\it e.g.}
specific heat and de Haas-van Alphen data \cite{Walm}.

The use of isoelectronic substitution  and its associated quantum criticality
is of particular interest in the context of superconductivity:
P substitution for As in the Ba 122 compound  destroys antiferromagnetism,
 the grey region in Fig.~6b and enables superconductivity,
  the brown region in the figure. Although direct access to the QCP
  is inhibited by the presence of superconductivity, it is observed
  that as $x$ is varied, the maximum $T_c(x)$ is found very
   close to $x=x_c$, similarly to what is observed in other
   strongly-correlated systems.
%%%%%%%%%%%%

\section{Superconductivity driven by antiferromagnetic correlations}

We have thus far
discussed
that electron correlations in the iron pncitides
and chalcogenides are sufficiently
strong to place these materials in the bad-metal regime,
and that frustrated magnetism of local moments
describes the dominant part of the magnetism.
What are the implications for unconventional superconductivity?
What are the structure and amplitude of the superconducting pairing?

The general picture for unconventional superconductivity was
 illustrated in Fig.~1b. Electron pairs
are formed through attractive interactions that are generated
while the Coulomb repulsion
is being avoided.
Pairing correlations will be disfavored between electrons on the same site,
which experience the dominant repulsive interactions.
Instead, pairing correlations develop among
electrons from different sites.
Because of this, the orbital part of the pairing wave function
tends to be in channels
orthogonal to the conventional $s$-wave channel;
these range from an extended $s$-wave to
cases with angular momentum larger than
$0$.

The proximity of the superconducting phase to the AF ordered phase
suggests the importance of the antiferromagnetic
correlations for superconductivity. Through the correlation effects
implicated by the bad-metal phenomenology
and the extensive information about the magnetic ordering and dynamics,
we have emphasized
the role of frustrating AF exchange interactions. The AF nature of such
 interactions favors
pairing to be in a spin-singlet channel, which is antisymmetric in spin space.
Because electrons are fermions, the overall wave function must be
antisymmetric under the exchange of
electrons. Consequently, the orbital part of the pairing wave function must be symmetric.

\subsection{Superconducting pairing structure and amplitude}

A characteristic feature of the iron pnictides and chalcogenides is that the
 bilinear exchange
 interactions contain both the nearest-neighbor interaction $J_1$
 and the next-nearest-neighbor interaction $J_2$.
In particular,
we have emphasized the microscopic reasons for the importance of $J_2$ interaction
(see Sec.~III.A),
which is also supported
by the fitting to the spin dynamics measured from neutron scattering experiments.
This $J_2$ interaction favors an extended $s$-wave pairing function,
with the leading term in the wave-vector dependence
being ${\rm cos} k_x {\rm cos}k_y$.
In relation to the tetrahedral  $D_{4h}$ point group symmetry, this belongs
to a pairing state with an $\mathrm{A}_{1g}$ symmetry.

%%% Figure 7%%%%
\begin{figure}[!b]
\begin{center}
\includegraphics[
%angle=270,
angle=0,
width=1.0\textwidth]{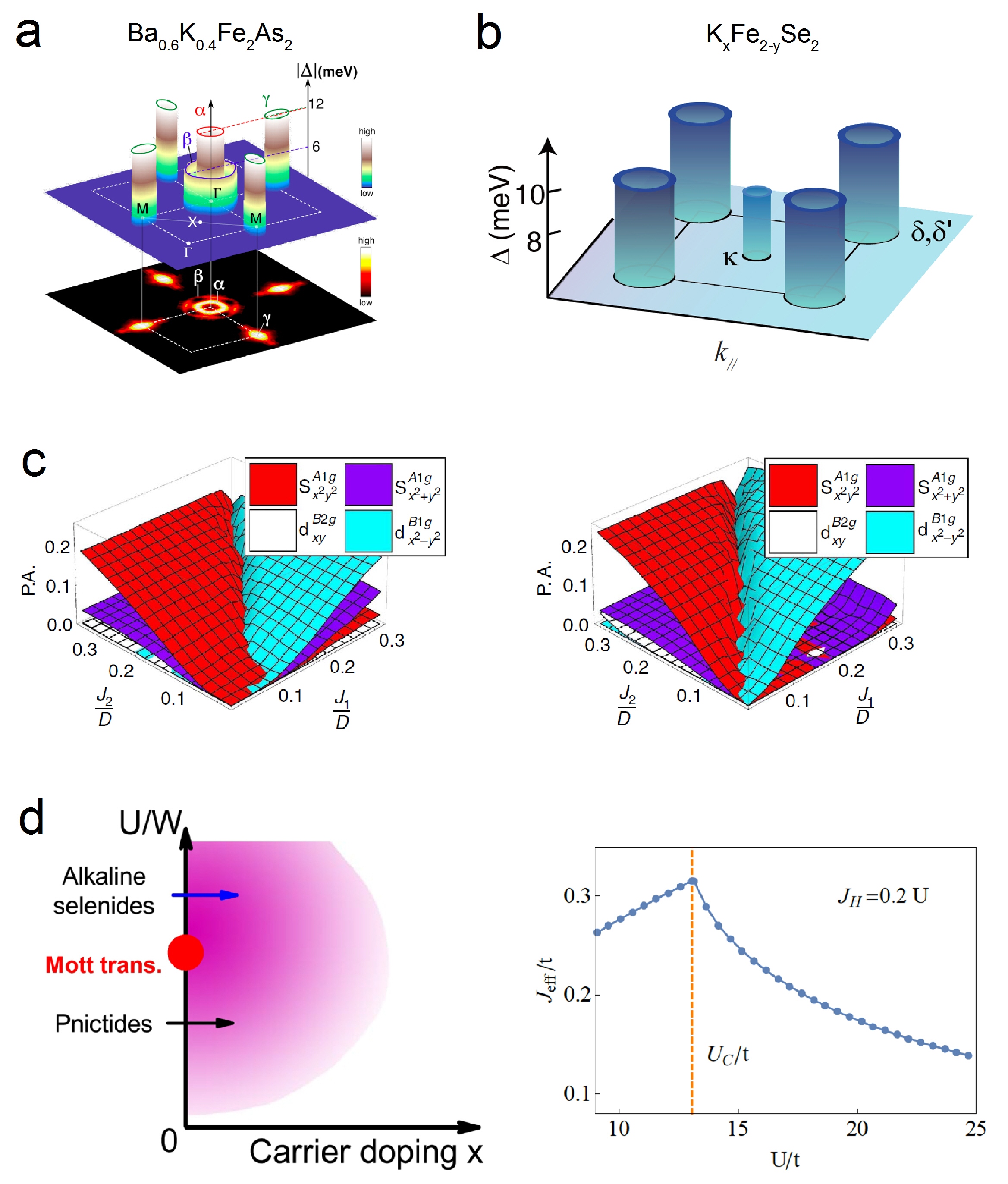}\\[-38ex]
\end{center}
\end{figure}
\newpage
\begin{figure}
\caption{
Superconductivity in the iron pnictides and chalcogenides.
Size of the superconducting gap $\Delta$:
a. on the hole ($\alpha$ and $\beta$) and electron ($\gamma$) Fermi pockets
 in Ba$_{0.6}$K$_{0.4}$Fe$_2$As$_2$.
From Ref.~\citenum{Ding.2008}; b. on the electron Fermi pockets of dominant
($\delta$ and $\delta^\prime$) and small
($\kappa$) spectral weight, in K$_x$Fe$_{2-y}$Se$_2$.
From Ref.~\citenum{XuFeng12}. c. Pairing amplitudes calculated
 for the iron pnictides and chalcogenides, from Ref.~\citenum{RYu.2013a}.
d. Left panel: Schematic phase diagram near a Mott transition. The purple shading
marks the parameter regime with strong antiferromagnetic correlations.
From Ref.~\citenum{RYu.2013a}. Right panel:
Exchange interaction plotted as a function of the Hubbard interaction $U$,
 for a Hund's coupling
$J_H=0.2 U$ in a two-orbital Hubbard model. From Ref.~\citenum{WDing.2014}.
}
\label{fig:7}
\end{figure}
%%%%%%%%%%%%

For
optimally doped iron pnictides, whose Fermi surfaces in the Brillouin zone typically
 consist of
the hole Fermi pockets near the $\Gamma=(0,0)$ point and
the electron Fermi pockets near $M=(\pm \pi, 0)$ and $(0,\pm \pi)$ points,
this extended $s$-wave
gap function is non-zero everywhere on the Fermi surfaces.
Moreover, it changes sign across the hole and electron pockets.
The pairing function is reflected in the single-particle excitation spectrum at the Fermi surface.
Where it is nonzero, a gap develops in the spectrum. Where it vanishes, the spectrum is gapless;
in other words, there is a node in the gap. ARPES measurements have found a spectrum
that is fully gapped. This is shown in Fig.~7a, where the gap function $\Delta$ is plotted
at the different parts of the Fermi surface; it is seen to have a magnitude of about
$10$ meV. The ratio $2 \Delta / T_c$ is about 4-7, thus
larger than the BCS value of $\sim 3.5$.

On the other hand, when $J_1$ is dominant, $d$-wave pairing
with a $(\cos k_x-\cos k_y)$ form factor and a ${\rm B}_{1g}$ symmetry is favored.
In the magnetic frustration regime, with $J_2$ and $J_1$ being comparable, a quasi-degeneracy
in pairing channels emerges in  model calculations.
It is interesting that such features arise from a variety of approaches, regardless of whether the
correlations are treated non-perturbatively or perturbatively
\cite{RYu.2013a, KSeo, Moreo.2009, WQChen.2009, DHLee.2013, Goswami.2010, Hirschfeld.2011, Graser.2009,Kuroki.2008,FWang.2009}.

To distinguish among the various approaches, it is instructive to compare the behavior of the gap
 in the iron chalcogenides,
especially since there are cases when they have only electron Fermi surfaces. This is illustrated
in Fig.~7b, for the case of the K$_x$Fe$_{2-y}$Se$_2$. We note that,
here too, the gap
is nodeless. Moreover, the magnitude of the gap for the electron Fermi pockets
having the dominant spectral weight (those near the $M$ points in the Brillouin zone)
is comparable with that of the iron pnictides.
This is consistent with the fact
that the superconducting transition temperature
($T_c$) of K$_x$Fe$_{2-y}$Se$_2$
is comparable to that of the iron pnictides.

The lesson of such comparisons is that the pairing amplitudes
are comparably large in the cases with or
 without  any hole pocket, that is to say with or without Fermi-surface nesting.
 This has a natural understanding in models where pairing
 is driven by short-range interactions such as the exchange interactions.
 Fig.~7c shows
 the pairing amplitude as a function of $J_1$ and $J_2$ in multi-orbital models
 with the pnictide (7c left) and chalcogenide (7c right) -type Fermi surfaces \cite{RYu.2013a}.
 This shows that,
 in spite of their distinctive Fermi surfaces, the dominant pairing channels as a function
 of the exchange interactions are similar in the two cases. Equally important, the pairing
 amplitudes are comparable in the two cases.

\subsection{Effects of orbital-selectivity on superconducting pairing}

As we have discussed in Sec.\ IIC,
electron correlations
can show strong
 orbital selectivity, which
 affects the electronic
 properties
 in the normal states significantly. A natural question is whether the pairing symmetry
 and amplitudes in the superconducting states
  are also
  influenced by the
  orbital selectivity.
  Studying this issue provides additional means to explore the effect of electron correlations
  on the properties of the superconducting state.

 Within the strong-coupling approach,
 the pairing order parameter is
defined in the orbital basis and the orbital selectivity can lead to some surprising effects.
In the usual case where the dominant pairing is in a full gap $s$-wave $A_{1g}$ channel,
sufficiently strong orbital selectivity may lead to
an anistropic gap
along the electron pockets and splitting
of spin resonance peaks \cite{Yu.2014b}.
Such effects have been
observed in experiments
by ARPES  \cite{Ge2013} and neutron scattering \cite{CZhang1,CZhang2},
respectively.

A more dramatic case arises in the context of the alkaline iron selenides.
As discussed in Sec.\ V A,
 in a multiorbital $J_1$-$J_2$ model with orbital independent
 $J_1$ and $J_2$, the dominant pairing symmetry is either an
$s$-wave $A_{1g}$ channel when $J_2$ is dominant,
or a $d$-wave $B_{1g}$ channel for dominant $J_1$ coupling.
A recent study~\cite{Nica.2015} on this model with orbital dependent $J_1$
and $J_2$ reveals that the orbital selectivity plays an essential role
in stabilizing an
intermediate
orbital-selective $B_{1g}$ pairing state that is constructed
 from the conventional $s$-wave $A_{1g}$ and $d$-wave $B_{1g}$
 states in the parameter regime where the two are quasi-degenerate.
  In this
  orbital-selective $B_{1g}$ pairing state, the dominant pairing amplitude
  is restricted to the $d_{xz}$ and $d_{yz}$ orbital subspace,
  in which the pairing state belongs to the $B_{1g}$ representation
  of the associated point group but has a form factor belonging to the
  $A_{1g}$ representation due to its nontrivial orbital structure.
  It is shown that for alkaline iron chalcogenides this pairing state has a full gap,
  and can produce a spin resonance at wave vector $(\pi,\pi/2)$,
  which is compatible with
   both the ARPES and neutron scattering
   experiments~\cite{Mou11,WangDing11,XuFeng12,WangDing12,ParkKeimer11,Friemel12}.

\subsection{Potential settings for optimizing superconductivity}

One grand challenge in physics is
whether there is a principle for optimizing superconducting $T_c$.
While superconductivity was initially observed in the iron pnictides
with nested hole and electron Fermi pockets,
we have already referred to the extensive recent
experiments that show
that iron chalcogenides with Fermi surface containing
only electron pockets exhibit superconductivity of comparable strength.
For example, single-layer FeSe on STO substrate has only electron Fermi pockets and, yet,
it holds the record of highest
$T_c$ in the FeSCs.
At the same time, Li(Fe,Co)As with almost perfectly nested Fermi pockets
 is non-superconducting~\cite{HDing.2015}.
 These new findings suggest that the Fermi surface geometry plays a secondary
 role in the superconductivity.

From a theoretical perspective,
the pairing amplitudes shown in Fig.~7c
for both iron pnictides and alkaline iron selenides
may shed some light on this important issue. The results indicate
that the pairing amplitudes are proportional to the ratio of $J/D$, where
$J$ is the antiferromagnetic exchange coupling and here and in Fig.~7c, $D$ is the effective bandwidth, 
as renormalized by electron-electron interactions,  whereas $W$ and $t$ in Fig.~7d are the bare bandwidth
 and kinetic-energy (``hopping") parameter, respectively.

This result can be generalized to the following principle,
which is illustrated in the left panel of Fig.~7d: the maximum of pairing
 amplitudes, and hence the optimal superconductivity is reached in the
  parameter regime of the phase diagram near the  Mott transition.
To see this, we note that, from a recent study on a multiorbital Hubbard model
 using a slave rotor approach, the exchange interactions increase as the
Mott transition is approached from both the insulating and the bad metal
sides (right panel of Fig. 7d).\cite{WDing.2014} At the same time, for a system with carrier doping,
the
renormalized bandwidth D is reduced as the Mott transition is
approached by decreasing doping. Correspondingly, at the
boundary between electronic localization and delocalization, the
ratio $J/D$ will be maximized and so will the superconducting
pairing amplitudes.
This principle suggests a route for optimizing superconductivity.
It also gives  an explanation of some key experimental
observations, such as the comparable $T_c$ in iron pnictides and alkaline iron selenides,
which have very different Fermi surface geometries.
In fact, experiments \cite{PCDai.2015}
revealed that these two materials have comparable
 exchange interactions.
In addition, both systems can be considered as being in proximity
to the point of Mott transition (Fig.~7d, left panel),
and therefore have a comparable degree of bad metallicity and renormalized bandwidth.

\section{Discussion}
The subject of the iron-based superconductors is so vast that it does not permit
a truly comprehensive survey of manageable length.
However, we wish to
discuss several additional issues that extend our discussions so far and
have the potential for much more in-depth studies.

\subsection{Quantum criticality and emergent phases beyond isoelectronic dopings}

The case for quantum criticality is quite strong near the optimal doping under
the isoelectronic P-for-As substitution in BaFe$_2$As$_2$, as we have discussed in
Sec.~\ref{sec:qcp}. By contrast, for carrier-doped iron arsenides, it remains to be
established whether the effects of quantum criticality operate  near optimal doping
over a substantial temperature
and energy range. The systems that have been investigated include
electron doping associated with Co- or Ni- substitution for Fe, and
hole doping induced by K-substitution for Ba.
Early evidence came from measurements of the thermopower
\cite{Gooch09}, nuclear magnetic resonance (NMR)~\cite{Ning10}
and ultrasonic attenuation~\cite{yoshizawa}.
Measurements based on the electrical transport \cite{nni08,jhchu09},
neutron scattering \cite{clester},
and  X-ray scattering \cite{,nandi} have
revealed the evolution of the transitions towards
an AF order as well as a
structural order (and the implicated electronic-nematic order)
in the Co-doped BaFe$_2$As$_2$ \cite{nandi}.
The corresponding transition temperatures $T_N$ and $T_s$ gradually
decrease with increasing $x$, which is suggestive of a QCP under
the superconducting dome.
Precisely how the lines, $T_N$ and $T_s$ {\it vs.} doping $x$ extrapolate 
towards the zero-temperature limit, however, has
been left open. Indeed,  NMR
experiments
have been interpreted in terms of two separate QCPs,
associated with the suppression of the AF and structural transitions, respectively\cite{Zheng.2013}.

This issue has been systematically investigated by
neutron and X-ray scattering studies of the
structural and magnetic phase transitions
in BaFe$_{2-x}$Ni$_x$As$_2$. \cite{XLu.2013}
The results are consistent with the
$T_N$, for the primary [$(\pi,0)$] AF order, and $T_s$ extrapolating towards
%a
the
 same QCP. However, before $T_N$ and $T_s$ can reach zero,
the transitions are interrupted by a new magnetic phase,
as illustrated  in Fig.~2b. The details of this magnetic phase are not
yet clear, although it shows glassy characteristics~\cite{Lu-prb2014}.
Still, the picture that has emerged is that, when the primary magnetic
order is sufficiently weakened, other emergent magnetic phases have
the chance to appear. In other words, quantum criticality operates over
a substantial dynamical range in the form of an ``avoided''
quantum criticality. The concurrent nature of the quantum criticality
associated with the primary AF order and the electronic nematic order
appears to have the same form as the quantum criticality
predicted for the iso-electronically tuned iron pnictides \cite{JDai.2009}.

Further studies are clearly called for to explore quantum
criticality in the carrier-doped iron arsenides. For example,
it would be illuminating to investigate the evolution of the low-temperature
electrical resistivity as a function of doping $x$ in the normal state
induced by suppressing superconductivity by a high magnetic field,
as was discussed in Sec.~\ref{sec:qcp}
 for the isoelectronic doping case \cite{Analytis.2014}.
This should be particularly promising, given that the electrical resistivity at optimal
Co-doping is quite close to being linear in $T$\, \cite{IFisher.2012}.
Likewise, ascertaining an $E/T$ scaling in the magnetic and nematic dynamics
by neutron scattering would be exceedingly instructive.

\subsection{Connection between iron pnictides and iron chalcogenides}

As we already discussed, iron pnictides were the focus of the attention
in the beginning of the field.
The main families of the iron pnictides, particularly the 1111 and 122 systems,
share many similarities in their electronic and magnetic structures:
the Fermi surfaces typically consist of nearly nested hole and electron pockets,
and the superconductivity is close to a $(\pi,0)$
collinear AF order
in the phase diagram.
Subsequently, the iron chalcogenides were extensively studied.
These systems show considerable variance in their electronic and magnetic
properties: some high $T_c$ compounds have only electron pockets, and the
AF ground state
could be considerably different from the $(\pi,0)$ AF order.

We have taken the perspective that most of the iron pnictides and essentially all
the chalcogenides have an electron occupancy on the $3d$ orbitals that is close
to $6$, and the degree of electronic correlations forms a continuum from the
iron phosphides through iron arsenides all the way to the iron chalcogenides.
A unified phase diagram is illustrated by the results of theoretical studies
on the correlation effects
-- of both the direct Coulomb repulsion $U$ and
the Hund's coupling $J_H$ --
 in multi-orbital Hubbard models,
similar to the one shown in the left panel of Fig.3c.\cite{Yu.2012,Yu2013,Yu-COSSMS2013}
This is consistent with the fact,
as already discussed in Sec.IIA,
that both iron pnictides and iron chalcogenides show bad metal behavior
in their normal states.
It is also compatible with the results of dynamical mean-field studies in a variety
of iron pnictides and chalcogenides \cite{Yin2011}.
Along this line of consideration, a physical pathway may be constructed to connect the insulating
245 alkaline iron selenides with their superconducting counterparts \cite{Yu2013,Gao2014},
providing the basis
for the placement of these materials in the
illustrative phase diagram shown in Fig.~7c, left panel.

What would be particularly instructive is to realize materials system in which the degree of electronic correlations, including
orbital selectivity, can be continuously tuned. Progress along this direction has
recently become possible. For example,
recent experiments have revealed the intriguing effect of
tuning the S-for-Se and Te-for-Se
substitution series  \cite{MYi-S-doping2015,Feng.2015} in the alkaline iron chalcogenides.
S-doping introduces negative chemical potential and leads to an enhanced kinetic energy,
effectively reducing the strength of electron correlations.
By contrast, Te-doping enhances correlation effects and moves the system towards
an insulating ground state.

\subsection{Extremely hole doped $A$Fe$_2$As$_2$ iron pnictides}
Yet another way to tune the system is to vary the 3$d$-electron filling  such that
it reaches far away from $N=6$.
The alkali iron pnictides $A$Fe$_2$As$_2$ with $A=$ K,
Rb, and Cs
provide an opportunity to study such effects. Valence counting gives $N=5.5$
 for these systems, corresponding to a large hole doping
relative to the usual parent iron arsenides.
This is verified by ARPES experiments, which
 show that the Fermi surfaces in these compounds
 contain only hole pockets~\cite{Sato09}.
 Even though their superconducting transition temperature $T_c$ is small ($3.5$ K in the case
  $A=$\,K, and becomes even smaller for the Rb and Cs cases),
  experimental evidence suggests that
 the superconductivity
 is unconventional~\cite{Reid12,Okazaki12,HongLi13,ZhangLi15}.
 Moreover,
there is a huge mass enhancement implying that they are strongly correlated
\cite{Hardy13,WangChen13,Grube2015}.

The fact that $A$Fe$_2$As$_2$ is in between the $N=6$ and $N=5$ limits
suggests that studying this system may link the properties of the usual systems
close to $N=6$
with those closer to $N=5$.
If all the five 3$d$ orbitals were degenerate,  $N=5$ would correspond
to half filling, while $N=6$ would be considerably away from half filling.
However, the 3$d$ orbitals in these systems are non-degenerate,
and we have discussed in
Sec.~\ref{subsec:osmp} that, even for $N=6$, the 3$d$ $xy$, $xz$ and $yz$ orbitals can be
close to half filling over an extended range of electron-electron interactions.
Studying the orbital selectivity as $N$ goes from $6$ to $5$ would therefore help reveal the
nature of the electron correlations in the iron pnictides \cite{Yu-COSSMS2013,deMedici2014}.

This issue is actively being studied experimentally.
There is
evidence that
a particular orbital (the $d_{xy}$ orbital, for example) is closer to Mott localization.
    In other words, the system may be closer to
  an orbital-selective
    Mott transition.
    The particularly enhanced correlations in the nearly localized
    orbital explain the unusually large Sommerfeld coefficients observed
    experimentally, while measurement of the Gr\"uneisen ratio suggests
    a proximity to quantum criticality as $A$ goes from K through Rb to Cs \cite{Grube2015}.
    It will be illuminating to understand the nature of this QCP and its detailed relationship
    to superconductivity, as well as to explore how to extend $N$ from $5.5$
    towards $5$ \cite{Grube2015}.

\section{Summary and outlook}
Concerted efforts
during the past seven years have led to a large materials basis
for superconductivity in the iron pnictides and chalcogenides.
The superconducting transition temperatures  are high
 compared to those of conventional superconductors
 and in the bulk materials at ambient pressure
are second only to those achieved in the copper oxides.
Moreover, recent experiments have shown that $T_c$ can be further increased
in the iron chalcogenides.

 The microscopic physics and superconductivity of the iron pnictides and chalcogenides
 is the subject of active ongoing studies, both theoretical and experimental. Still, extensive understandings
 have already been achieved, which we have summarized. These include:
\begin{itemize}
\item The normal state of the iron-based superconductors is a bad metal. More specifically,
the DC electrical resistivity
at room temperature is large, reaching the Mott-Ioffe-Regel limit.
Relatedly, the
Drude weight of the optical conductivity is strongly
suppressed compared to its non-interacting counterpart.
A relatively modest enhancement of the electron-correlation strength through a
kinetic-energy reduction leads to a fully localized, Mott-insulating regime.
These and other properties
provide compelling evidence that the iron pnictides and chalcogenides are strongly
correlated in that the effect of electron correlations is non-perturbative. They also
provide evidence for the normal state
being
in proximity to a Mott transition.
Relatedly, orbital selectivity has been
 found in the bad-metal regime,
with its extreme form being an orbital-selective Mott phase.

\item The spin excitations map out a spin-wave-like
spectrum.
The total spectral weight of the
observed excitations is large, in that it corresponds to a local moment on the order of
one to several $\mu_B$ per Fe. Across the materials basis, the variation in the size
of the local moments goes well
with the variation in the degree of electron correlations inferred from electrical transport
and charge
dynamics studies. The energy dispersions of the spin excitations implicate a $J_1$-$J_2$ magnetic
frustration, as was theoretically recognized from the beginning of the field.
The detailed energy and momentum distributions of the spin spectral weight
encode the
damping of the spin excitations,
whose variation
 across the materials basis
 also seems consistent with
the corresponding variation in the degree of electron correlations.

\item Electronic nematicity has been observed in the normal state of many if not all the FeSCs.
There is considerable evidence that its origin lies in
the spin degrees of freedom,
although the issue remains to be fully settled.

\item Compelling experimental evidence has accumulated for quantum criticality
predicted to occur in the isoelecronic P-doped iron arsenides. There is also indication
 for quantum criticality in the carrier-doped iron arsenides, although the case here needs to
 be further substantiated.

\item Superconductivity is unconventional in that it is primarily driven
by electron-electron interactions and, relatedly, that the superconducting pairing
is not in the conventional $s$-wave channel.
We have summarized the arguments for superconductivity being primarily driven
by antiferromagnetic correlations, and discussed the implications of the comparably
large $T_c$ and pairing amplitudes observed in the FeSCs with
or without a Fermi-surface nesting. Finally, there is considerable theoretical basis
for the quasi-degeneracy among several competing channels of superconducting pairing.
\end{itemize}

The properties observed in these materials and the theoretical considerations
outlined here suggest that optimized superconductivity occurs in bad metals
that are not only close to a magnetic order but also at the boundary
 of electronic localization and delocalization.
This notion connects the FeSCs
 to unconventional
 superconductivity that has been observed in other classes of strongly correlated materials, such
 as the copper oxides, heavy fermion metals and organic charge-transfer salts.
 Indeed, the normal state of all these other superconductors
 satisfies  the criterion of bad metals.
Therefore,
higher $T_c$ superconductivity may appear in
materials which possess even larger antiferromagnetic exchange
interactions but retain bad metallicity.

\vspace{0.5cm} \noindent{\bf Acknowledgments}
We would like to thank J. Analytis, M. Bendele, P. C. Dai, W. Ding, L. Harriger, 
X. Lu and P. Nikolic
for their helpful input on the manuscript.
We have benefited from collaborations and/or discussions with
J. Dai, P. C. Dai, W. Ding, P. Goswami, K. Grube, D. H. Lu, A. H. Nevidomskyy,
E. Nica, P. Nikolic, Z.-X. Shen, H. von L\"ohneysen, Z. Wang, M. Yi, and J.-X. Zhu.
The work has been supported in part  by the NSF Grant No.\ DMR-1309531 and
the Robert A.\ Welch Foundation Grant No.\ C-1411 (Q.S.),
and by the  National  Science
Foundation of China Grant number 11374361
and the Fundamental Research Funds for the Central Universities and the
Research Funds of Renmin University of China (R.Y.).
We acknowledge the support provided in part by the
NSF Grant  No. NSF PHY11-25915 at KITP, UCSB, for our participation
in the Fall 2014 program on ``Magnetism, Bad Metals
and Superconductivity: Iron Pnictides and Beyond". Q.S. and E.A.
also acknowledge the hospitality of the Aspen Center for Physics
(NSF Grant No.\ 1066293).

%\vspace{0.5cm}
%\noindent{\bf Competing interests statement} The authors declare that
%they have no competing financial interests.


\begin{thebibliography}{99}

\bibitem{Hos2008}
Kamihara, Y., Watanabe, T., Hirano, M. \& Hosono, H.
Iron-Based Layered Superconductor La[O$_{1-x}$F$_x$]FeAs (x = 0.05-0.12) with $T_c$ = 26 K.
{\it J. Am. Chem. Soc.} {\bf 130}, 3296 (2008).

\bibitem{Bednorz-Mueller1986}
Bednorz, J. G., M\"{u}ller, K. A. Possible High$T_c$ Superconductivity in the Ba-
La-Cu-O System. {\it Z. Phys.\ B- Condensed Matter} {\bf 64}, 189 (1986).

\bibitem{ChenRen2008}
Ren Z. A. {\it et al.}, Superconductivity at 55K in Iron-Based F-Doped Layered Quaternary Compound Sm[O$_{1-x}$F$_x$]FeAs. {\it Chin. Phys. Lett.} {\bf 25}, 2215 (2008).

\bibitem{Xue.2012}
Wang, Q.-Y. {\it et al.}
Interface-Induced High-Temperature Superconductivity in Single Unit-Cell FeSe Films on SrTiO$_3$.
{\it Chin. Phys. Lett.} {\bf 29}, 037402 (2012).

\bibitem{SLHe2013}
He, S. {\it et al.} Phase Diagram and High Temperature Superconductivity at 65 K in Tuning Carrier Concentration of Single-Layer FeSe Films. {\it Nat. Mater.} {\bf 12}, 605 (2013).

\bibitem{Shen.2014}
Lee,  J. J. {\it et al.}
Interfacial mode coupling as the origin of the enhancement of T$_c$ in FeSe films on SrTiO$_3$.
{\it Nature} {\bf 515}, 245?48 (2014).

\bibitem{YWang2015}
Zhang, Z. {\it et al.}
Onset of the Meissner effect at 65 K in FeSe thin film grown on Nb doped SrTiO3 substrate.
{\it Science Bulletin} {\bf 60}, 1301-1304 (2015).

\bibitem{Jia.2014}
Ge, J.-F., {\it et al.},
Superconductivity above 100 K in single-layer FeSe films on doped SrTiO$_3$.
{\it Nat. Mater.} {\bf 14}, 285 (2015).

\bibitem{Onnes}Kamerlingh Onnes, H.
{\it Commun. Phys. Lab. Univ. Leiden}. Suppl. {\bf 29} (1911).

\bibitem{BCS} Bardeen, J.,Cooper, L. \& Schrieffer, J.R.
Microscopic Theory of Superconductivity. {\it Phys. Rev.} {\bf 106}, 162 (1957).

\bibitem{Boeri2008}
Boeri, L., Dolgov, O. V. \& Golubov, A. A.
Is LaFeAsO$_{\rm 1-x}$F$_{\rm x}$ an Electron-Phonon Superconductor?
 {\it Phys.~Rev.~Lett.}  {\bf 101}, 026403 (2008).

\bibitem{SiAbrahams.2008}Si, Q. \& Abrahams, E.
Strong Correlations and Magnetic Frustration in the High $T_c$ Iron Pnictides.
{\it Phys. Rev. Lett.} {\bf 101}, 076401 (2008)

\bibitem{Basov.2009}
Qazilbash, M.M. {\it et al}
Electronic correlations in the iron pnictides
{\it Nature Physics} {\bf 5}. 647 (2009).

\bibitem{Yin2011}
Yin, Z. P., Haule, K. \& Kotliar, G.
Kinetic frustration and the nature of the magnetic and paramagnetic states
in iron pnictides and iron chalcogenides.
{\it Nature Mater.} {\bf 10}, 932 (2011).

\bibitem{PCDai.2015}
Dai, P. C. Antiferromagnetic order and spin dynamics in iron-based superconductors.
{\it Rev. Mod. Phys.} {\bf 87}, 855 (2015).

\bibitem{IFisher.2012}
Chu,  J. H., Kuo, H.-H., Analytis, J. G.  \& Fisher, I. R.
Divergent Nematic Susceptibility in an Iron Arsenide Superconductor.
{\it Science} {\bf 337}, 710-712 (2012).

\bibitem{HHKuo2015}
Kuo, H.-H., Chu, J. -H., Kivelson, S. A. \& Fisher, I. R.
Ubiquitous signatures of nematic quantum criticality in optimally doped
Fe-based superconductors.
{\it arXiv}:1503.00402v1.

\bibitem{JDai.2009}
J. Dai, Q. Si, J.-X. Zhu, and E. Abrahams,
Iron pnictides as a new setting for quantum criticality.
{\it Proc. Natl. Acad. Sci. USA} {\bf 106}, 4118 - 4121 (2009).

\bibitem{Matsuda.2012}
Hashimoto, K. {\it et al.} A Sharp Peak of the Zero-Temperature Penetration Depth at Optimal Composition in BaFe$_2$(As$_{1-x}$P$_x$)$_2$. {\it Science} {\bf 336}, 1554 (2012).

\bibitem{Analytis.2014}
Analytis, J.G. {\it et al.}
Transport near a quantum critical point in BaFe$_2$(As$_{1-x}$P$_x$)$_2$. {\it Nature Phys.} {\bf 10}, 194 (2014).

\bibitem{RYu.2013a}
Yu, R., Goswami, P., Si, Q. Nikolic, P., and Zhu, J.-X.
Superconductivity at the Border of Electron Localization and Itinerancy. {\it Nat. Commun.} {\bf 4}, 2783 (2013).

\bibitem{Hirschfeld.2011}
Hirschfeld, P. J., Korshunov, M. M. and Mazin, I. I. Gap symmetry and structure of Fe-based superconductors. {\it Rep. Prog. Phys.} {\bf 74}, 124508 (2011).

\bibitem{Rotter.2008}
Rotter, M., Tegel, M. \& Johrendt, D.
Superconductivity at 38 K in the Iron Arsenide
(Ba$_{1-x}$K$_x$)Fe$_2$As$_2$.
{\it Phys. Rev. Lett.} {\bf 101}, 107006 (2008).


\bibitem{HsuWu2008} Hsu, F.-C. {\it et al.} Superconductivity in the PbO-type structure $\alpha$-FeSe. {\it Proc. Natl. Acad. Sci. USA} {\bf 105}, 14262 (2008).

\bibitem{JGuo.2010}
Guo, J. \textit{et al}. Superconductivity in the iron selenide K$_x$Fe$_2$Se$_2$
($0\leqslant x\leqslant1.0$). \textit{Phys. Rev. B} \textbf{82}, 180520 (2010).

\bibitem{LSun.2012}
Sun, L. \textit{et al}.
Re-emerging superconductivity at $48$ Kelvin in iron chalcogenides.
\textit{Nature} \textbf{483}, 67-69 (2012).

\bibitem{XLu.2013}
Lu X. {\it et al.} Avoided quantum criticality and magnetoelastic coupling
in BaFe$_{2-x}$Ni$_{x}$As$_{2}$. {\it Phys. Rev. Lett.} {\bf 110}, 257001 (2013).

\bibitem{Yi09}
Yi, M. {\it et al.} Electronic Structure of the BaFe$_2$As$_2$ Family
of Iron Pnictides. {\it Phys. Rev. B} {\bf 80}, 024515 (2009).

\bibitem{Sato09} Sato, T. {\it et al.} Band Structure and Fermi Surface
of an Extremely Overdoped Iron-Based Superconductor KFe$_2$As$_2$.
{\it Phys. Rev. Lett.} {\bf 103}, 047002 (2009).

\bibitem{Hussey2004}
 Hussey, N. E., Takenaka, K., and Takagi, H.
 Universality of the Mott-Ioffe-Regel limit in metals.
{\it Philoso. Mag.} {\bf 84}, 2847 (2004).

\bibitem {Fawcett-RMP}
Fawcett, E.
Spin-density-wave antiferromagnetism in chromium
{\it Rev. Mod. Phys.} {\bf 60}, 209( 1988).

\bibitem{Si-NJP2009}
Si, Q., Abrahams, E., Dai, J. \& Zhu, J.-X.
Correlation effects in the iron pnictides.
{\it New Journal of Physics} {\bf 11}, 045001 (2009).

 \bibitem{Nakajima.2014}
 M. Nakajima {\it et al.}
 Normal-state charge dynamics in doped BaFe$_2$As$_2$: Roles of doping and necessary
 ingredients for superconductivity.
{\it Sci. Rep.} {\bf 4}, 5873 (2014).

\bibitem{Tamai10} Tamai, A. {\it  et al.}
Strong electron correlations in the normal state of the iron-based FeSe$_{0:42}$Te$_{0:58}$
superconductor observed by angle-resolved photoemission spectroscopy.
 {\it Phys. Rev. Lett.} {\bf 104}, 097002 (2010).

\bibitem{MYi.2013}
Yi, M. {\it et al.}
Observation of temperature-induced crossover to an orbital-selective Mott
phase in
${\rm A_xFe_{2-y}Se_2}$~(A=K, Rb) superconductors.
{\it Phys.~Rev.~Lett.} {\bf 110}, 067003 (2013).

\bibitem{MYi.2015}
Yi, M. {\it et al.}
Observation of universal strong orbital-dependent correlation effects
in iron chalcogenides.
{\it Nature Commun.} {\bf 6}, 7777 (2015).

\bibitem{Si-NatPhys.2009} Si, Q. Iron pnictide superconductors: Electrons on the verge. {\it Nat. Phys.} {\bf 5}, 629 (2009).

\bibitem{Zhu2010}
 Zhu, J.-X. {\it et al.}
Band narrowing and Mott localization in iron oxychalcogenides
La$_2$O$_2$Fe$_2$O(Se,S)$_2$. {\it Phys.~Rev.~Lett.}
 {\bf 104}, 216405 (2010).

 \bibitem{Fang2011}
 Fang, M.-H. {\it et al.}
Fe-based superconductivity with $T_c=31$ K bordering an antiferromagnetic
insulator in (Tl,K)Fe$_x$Se$_2$.
{\it Europhys. Lett.} {\bf 94}, 27009 (2011).

 \bibitem{Yu2013}
 Yu R. \& Si Q.
Orbital-selective Mott phase in multiorbital models for alkaline iron
selenides K$_{\rm 1-x}$Fe$_{\rm 2-y}$Se$_{\rm 2}$.
{\it Phys.~Rev.~Lett.} {\bf 110}, 146402 (2013).

\bibitem{neutrons}
Free, D. G. \& Evans, J. S. O. Low-temperature nuclear and magnetic structures
of La$_2$O$_2$Fe$_2$OSe$_2$ from x-ray and neutron diffraction measurements.
{\it Phys. Rev. B} {\bf 81}, 214433 (2010).

\bibitem{Imada-RMP}
Imada, M., Fujimori, A. \& Tokura, Y.
Metal-insulator transitions.
{\it Rev. Mod. Phys.} {\bf 70}, 1039 (1998).

\bibitem{Freelon}
Freelon, B. {\it et al.}
Mott-Kondo Insulator Behavior in the Iron Oxychalcogenides.
arXiv:1501.00332.

\bibitem{Bao.2011} Bao, W. {\it et al.} A Novel Large Moment Antiferromagnetic
Order in K$_{0.8}$Fe$_{1.6}$Se$_2$ Superconductor. {\it Chin. Phys. Lett.}
{\bf 28}, 086104 (2011).

\bibitem{Yu2011}
Yu, R., Zhu, J.-X. \& Si, Q.
Mott transition in Modulated Lattices and Parent Insulator of ${\rm (K,Tl)_yFe_xSe2}$ Superconductors.
{\it Phys. Rev. Lett.} {\bf 106}, 186401 (2011).

\bibitem{Zhou2011}
Zhou, Y., Xu, D.-H., Zhang, F.-C. \& Chen, W.-Q.
Theory for superconductivity in (Tl,K)Fe$_x$Se$_2$ as a doped Mott insulator. {\it EuroPhys. Lett.} {\bf 95}, 17003 (2011).

\bibitem{Birgeneau2015}  Wang, M. {\it et al.}
Mott localization in a pure stripe antiferromagnet Rb$_{1-\delta}$Fe$_{1.5-\sigma}$S$_2$.
{\it Phys. Rev. B} {\bf 92}, 121101 (2015)

 \bibitem{Wang2014}
 Wang, Z. {\it et al.}
Orbital-selective metal-insulator transition and gap formation above
$T_c$ in superconducting Rb(1-x)Fe(2-y)Se$_2$.
{\it Nat.~Commun.}
{\bf 5}, 3202 (2014).

\bibitem{Ding2014}
Ding, X., Pan, Y., Yang, H. \& Wen, H.-H.
Strong and nonmonotonic temperature dependence of Hall coefficient in
superconducting K$_x$Fe$_{2-y}$Se$_2$ single crystals.
{\it Phys.~Rev. B} {\bf 89}, 224515 (2014).

\bibitem{Li2014}
Li, W. {\it et al.}
Mott behaviour in K$_x$Fe$_{2-y}$Se$_2$ superconductors studied by pump-probe
spectroscopy.
{\it Phys.~Rev.~B} {\bf 89}, 134515 (2014).

\bibitem{Gao2014}
Gao, P. {\it et al.}
Role of the 245 phase in alkaline iron selenide superconductors
revealed by high-pressure studies.
{\it Phys.~Rev.~B}
{\bf 89}, 094514 (2014).

\bibitem{Yu-COSSMS2013}
Yu R., Zhu J.-X. \& Si Q.
Orbital-dependent effects of electron correlations in microscopic models
for iron-based superconductors.
{\it Current Opinion in Solid State and Materials Science}
 {\bf 17}, 65-71 (2013).

\bibitem {Anisimov2002}
Anisimov, V. I., Nekrasov, I. A., Kondakov, D. E., Rice, T. M.  \& Sigrist, M.
Orbital-selective Mott-insulator transition in Ca$_{2-x}$Sr$_x$RuO$_4$.
{\it Eur. Phys. J. B} {\bf 25}, 191 (2002).

\bibitem{Yu.2012}
Yu, R. \& Si, Q.
$U(1)$ Slave-spin theory and its application to Mott transition
in a multi-orbital model for iron pnictides.
{\it Phys. Rev. B} {\bf 86}, 085104 (2012)

\bibitem{deMedici2014} de'Medici, L., Giovannetti, G. \& Capone, M. Selective Mottness
as a key to iron superconductors. {\it Phys. Rev. Lett.} {\bf 112}, 177001 (2014).

\bibitem{Bascones}
Bascones, E., Valenzuela, B. \& Calder\'{o}n, M. J. Orbital differentiation and the role
of orbital ordering in the magnetic state of Fe superconductors.
{\it Phys. Rev. B} {\bf 86}, 174508 (2012).

\bibitem{Fang08}
Fang, C., Yao, H., Tsai, W.-F., Hu, J. P., and Kivelson, S. A.
Theory of electron nematic order in LaFeAsO,
{\it Phys. Rev. B} {\bf 77}, 224509 (2008).

\bibitem{Xu08} Xu, C., Muller, M., and Sachdev, S.
Ising and spin orders in the iron-based superconductors,
{\it Phys. Rev. B} {\bf 78}, 020501(R) (2008).

\bibitem{Laad.2009}Laad, M. S., Craco, L., Leoni. S. \&Rosner, H.
Electrodynamic response of incoherent metals: Normal phase of iron pnictides.
{\it Phys. Rev. B} {\bf 79} 024515 (2009).

\bibitem{KSeo} Seo, K., Bernevig, B. A. \& Hu, J.
Pairing Symmetry in a Two-Orbital Exchange Coupling Model of Oxypnictides.
Phys. Rev. Lett. \textbf{101}, 206404 (2008).

\bibitem{Moreo.2009} Moreo, A., Daghofer, M., Riera, J. A. \& Dagotto, E.
Properties of a two-orbital model for oxypnictide superconductors: Magnetic order,
$B_{2g}$ spin-singlet pairing channel, and its nodal structure
{\it Phys. Rev. B} \textbf{79}, 134502 (2009).

\bibitem{WQChen.2009}
Chen, W.-Q., Yang, K.-Y., Zhou, Y., and Zhang, F.-C.
Strong Coupling Theory for Superconducting Iron Pnictides.
{\it Phys. Rev. Lett.} {\bf 102}, 047006 (2009).

\bibitem{DHLee.2013} Yang, F., Wang, F., and Lee, D.-H.
Fermiology, orbital order, orbital fluctuations, and Cooper pairing in iron-based superconductors.
{\it Phys. Rev. B} {\bf 88}, 100504 (R) (2013).

\bibitem{Berg.2009} Berg, E., Kivelson, S. A. \& Scalapino, D. J. A twisted ladder:
relating the Fe superconductors to the high-$T_c$ cuprates.
\textit{New J. Phys.} \textbf{11}, 085007 (2009).

\bibitem{Lv} Lv, W., Kr\"{u}ger, F., and Phillips, P. Orbital ordering and
unfrustrated ($\pi$,0) magnetism from degenerate double exchange in the iron pnictides.
{\it Phys. Rev. B} {\bf 82}, 045125 (2010).

\bibitem{Yu2015}
Yu, R. \& Si, Q.
Antiferroquadrupolar and Ising-nematic orders of a frustrated
bilinear-biquadratic Heisenberg model and implications
for the magnetism of FeSe.
{\it Phys.~Rev.~Lett.} {\bf 115}, 116401 (2015).

\bibitem{FWang2015}
Wang, F., Kivelson, S. A. \& Lee, D.-H.
Nematicity and quantum paramagnetism in FeSe.
{\it Nature Phys.} (2015); online at doi:10.1038\/nphys3456.

\bibitem{Cruz2008}
de la Cruz, C. {\it et al.} Magnetic Order versus superconductivity in the Iron-based layered La(O$_{1-x}$F$_x$)FeAs systems. {\it Nature} {\bf 453}, 899 (2008).

\bibitem{Diallo10}
Diallo, S. O. {\it et al.} Paramagnetic spin correlations in CaFe$_2$As$_2$ single crystals. {\it Phys. Rev. B} \textbf{81}, 214407 (2010).

\bibitem{Harriger11}
Harriger, L. W. {\it et al.} Nematic spin fluid in the tetragonal phase of BaFe$_2$As$_2$. {\it Phys. Rev. B} {\bf 84}, 054544 (2011).

\bibitem{Ewings11}
Ewings, R. A. {\it et al.} Itinerant spin excitations in SrFe$_{2}$As$_{2}$ measured by inelastic neutron scattering. {\it Phys. Rev. B} {\bf 83}, 214519 (2011).


\bibitem{Yildirim.2008}
Yildirim, T.
Origin of the 150-K Anomaly in LaFeAsO: Competing antiferromagnetic
interactions, frustration, and a structural phase transition.
{\it Phys. Rev. Lett.} {\bf 101}, 057010 (2008).

\bibitem{Ma2008}
Ma, F., Lu, Z.-Y. \& Xiang, T.
Antiferromagnetic superexchange interactions in LaOFeAs.
{\it Phys. Rev. B} {\bf 78}, 224517 (2008).

\bibitem{Chandra90}
Chandra, P., Coleman, P. \& Larkin, A. I.
Ising transition in frustrated Heisenberg models.
{\it Phys. Rev. Lett.} {\bf 64}, 88-91 (1990).

\bibitem {Fazekas}Fazekas, P.
{\it Lecture Notes on Electron Correlation and Magnetism}
(World Scientific, Singapore, 1999), Chap. 5.

\bibitem {JJK}Yu, R. {\it et al.}
Spin dynamics of a $J_1-J_2-K$ model for the paramagnetic phase of iron pnictides.
{\it Phys. Rev. B} {\bf 86}, 085148 (2012)

\bibitem{Wysochi.2011}
Wysocki, A. L., Belashchenko, K. D. \& Antropov, V. P.
Consistent model of magnetism in ferropnictides.
{\it Nature Phys.} 7, 485 (2011).

\bibitem{Liu.2012} Liu, M. S. {\it et al.} Nature of magnetic excitations in superconducting
BaFe$_{1.9}$Ni$_{0.1}$As$_{2}$. {\it Nat. Phys.} {\bf 8}, 376-381 (2012).

\bibitem{Dong08} Dong, J. {\it et al.} Competing orders and spin-density-wave instability
 in La(O$_{1-x}$F$_x$)FeAs. {\it Europhys. Lett.} {\bf 83}, 27006 (2008).

\bibitem{Knolle11} Knolle, J., Eremin, I. \& Moessner, R. Multiorbital spin susceptibility
in a magnetically ordered state: Orbital versus excitonic spin density wave scenario.
{\it Phys. Rev. B} {\bf 83}, 224503 (2011).

\bibitem{Ma2009} Ma, F., Ji, W., Hu, J., Lu, Z.-Y. and Xiang, T. First-Principles Calculations
of the Electronic Structure of Tetragonal $\alpha$-FeTe and $\alpha$-FeSe Crystals:
Evidence for a Bicollinear Antiferromagnetic Order. {\it Phys. Rev. Lett.}
{\bf 102}, 177003 (2009).

\bibitem{Wen2015}
Wen, J. Magnetic neutron scattering studies on the Fe-based superconductor system
Fe$_{1+y}$Te$_{1-x}$Se$_x$. {\it Annals Phys.} {\bf 358}, 92 (2015).

\bibitem{MWang2011} Wang, M. {\it et al.} Spin waves and magnetic exchange interactions in insulating Rb$_{0.89}$Fe$_{1.58}$Se$_2$. {\it Nat. Commun.} {\bf 2}, 580 (2011).

\bibitem{Terashima15} Terashima, T. {\it et al.} Pressure-Induced Antiferromagnetic Transition and Phase Diagram in FeSe. {\it J. Phys. Soc. Jpn.} {\bf 84}, 063701 (2015).

\bibitem{Bendele12} Bendele, M. {\it et al.} Coexistence of superconductivity and magnetism in FeSe$_{1-x}$ under pressure. {\it Phys. Rev. B} {\it 85}, 064517 (2012).

\bibitem{Yu-Goswami-Si2011}
Yu, R., Goswami, P. \& Si, Q. The magnetic phase diagram of an extended $J_1$-$J_2$ model
on a modulated square lattice and its implications for the antiferromagnetic phase
of K$_y$Fe$_x$Se$_2$. {\it Phys. Rev. B} {\bf 84}, 094451 (2011).

\bibitem{Cao-Dai2011}
Cao, C. \& Dai, J. Block Spin Ground State and Three-Dimensionality
of (K,Tl)Fe$_{1.6}$Se$_2$. {\it Phys. Rev. Lett.} {\bf 107}, 056401 (2011).

\bibitem{Chi13} Chi, S. {\it et al.} Neutron scattering study of spin dynamics in superconducting (Tl,Rb)$_2$Fe$_4$Se$_5$. {\it Phys. Rev. B} {\bf 87}, 100501 (2013).

\bibitem{MWang2013} Wang, M. {\it et al.} Two spatially separated phases in semiconducting
Rb$_{0.8}$Fe$_{1.5}$S$_2$. {\it Phys. Rev. B} {\bf 90}, 125148 (2014).

\bibitem{MYi.2011} Yi, M. {\it et al.} Symmetry breaking orbital anisotropy on detwinned
Ba(Fe$_{1-x}$Co$_x$)$_2$As$_2$ above the spin density wave transition. {\it Proc. Natl. Acad. Sci. USA} {\bf 108}, 6878 (2011).

\bibitem{Fernandes14} Fernandes, R. M., Chubukov, A. V., and Schmalian, J. Nematic order in iron superconductors - who is in the driver's seat? {\it Nat. Phys.} {\bf 10}, 97 (2014).

\bibitem{Devereaux10} Chen, C.-C. {\it et al.}
Orbital Order and Spontaneous Orthorhombicity in Iron Pnictides.
{\it Phys. Rev. B} {\bf 82}, 100504(R) (2010).

\bibitem{Lee2009}
Lee, C. C., Yin, W. G. \& Ku, W.
Ferro-Orbital Order and Strong Magnetic Anisotropy in the Parent Compounds
of Iron-Pnictide Superconductors.
{\it Phys. Rev. Lett.} {\bf 103}, 267001 (2009).

\bibitem{Kruger2009} Kr\"{u}ger, F., Kumar, S., Zaanen, J., and van den Brink, J. Spin-orbital frustrations and anomalous metallic state in iron-pnictide superconductors. {\it Phys. Rev. B} {\bf 79}, 054504 (2009).

\bibitem{Lu-Science2014}
Lu, X. {\it et al.}
Nematic spin correlations in the tetragonal state of uniaxial strained
BaFe$_{2-x}$Ni$_x$As$_2$.
{\it Science} {\bf 345}, 657-660 (2014).

\bibitem{McQueen09} McQueen, T. M. {\it et al.}
Tetragonal-to-Orthorhombic Structural Phase Transition at 90 K in the Superconductor Fe$_{1.01}$Se.
{\it Phys. Rev. Lett.} {\bf 103}, 057002 (2009).

\bibitem{Medvedev09} Medvedev, S. {\it et al.}
Electronic and magnetic phase diagram of $\beta$-Fe$_{1.01}$Se with superconductivity
at 36.7 K under pressure.
{\it Nat. Mater.} {\bf 8}, 630 (2009).

\bibitem{Bohmer14} B\"{o}hmer, A. E. {\it et al.}
Origin of the Tetragonal-to-Orthorhombic Phase Transition in FeSe: A Combined Thermodynamic and NMR Study of Nematicity.
{\it Phys. Rev. Lett.} {\bf 114}, 027001 (2015).

\bibitem{Baek14} Baek, S.-H. {\it et al.}
Orbital-driven nematicity in FeSe.
{\it Nat. Mater.} {\bf 14}, 210 (2015).

\bibitem{Glasbrenner2015}
Glasbrenner J. K. {\it et al.}
Effect of magnetic frustration on nematicity and superconductivity in iron chalcogenides.
{\it Nature Phys.} (2015); online at doi:10.1038\/nphys3434.

\bibitem{Rahn15}
Rahn, M. C., Ewings, R. A., Sedlmaier, S. J., Clarke, S. J. \& A. T. Boothroyd,
Strong $(\pi,0)$ spin fluctuations in $\beta$-FeSe observed by neutron
spectroscopy.
{\it Phys. Rev. B} {\bf 91}, 180501(R) (2015).

\bibitem{WangZhao15}
Wang, Q. {\it et al.}
Strong interplay between stripe spin fluctuations, nematicity
and superconductivity in FeSe'', arXiv:1502.07544.

\bibitem {AS} Abrahams, E. \& Si, Q.
Quantum criticality in the iron pnictides and chalcogenides.
{\it J. Phys.: Condens. Matter} {\bf 23}, 223201 (2011).

 \bibitem {YMat}
 Kasahara, S. {\it et al.}
 Evolution from non-Fermi- to Fermi-liquid transport via isovalent doping
in BaFe$_2$(As$_{1-x}$P$_x$)$_2$ superconductors.
 {\it Phys.~Rev.~B} {\bf 81}, 184519 (2010).


  \bibitem{LRVW}L{\" o}hneysen, H., Rosch, A., Vojta, M., \& W{\" o}lfle, P.
  Fermi-liquid instabilities at magnetic quantum phase transitions. {\it Revs. Modern Phys.} {\bf 79}, 1015 (2007).

  \bibitem{KW}Kadowaki, K. \& Woods, S.
  Universal relationship of the resistivity and specific heat in heavy-fermion compounds. {\it Solid State Commun.} {\bf 58}, 507-509 (1986).

  \bibitem {Walm} Walmsley, P. {\it et al.}
  Quasiparticle Mass Enhancement Close to the Quantum Critical Point in BaFe$_2$(As$_{1-x}$P$_x$)$_2$. {\it Phy. Rev. Lett.} {\bf 110}, 257002 (2013).

\bibitem{Ding.2008} Ding, H. {\it et al.} Observation of Fermi-surface-dependent nodeless superconducting gaps in Ba$_{0.6}$K$_{0.4}$Fe$_2$As$_2$.
{\it Europhys. Lett.} {\bf 83}, 47001 (2008).

\bibitem{XuFeng12} Xu, M. {\it et al.} Evidence for an $s$-wave
superconducting gap in K$_x$Fe$_{2-y}$Se$_2$ from angle-resolved
photoemission. {\it Phys. Rev. B} {\bf 85}, 220504 (2012).

\bibitem{WDing.2014} Ding, W., Yu, R., Si, Q. \&Abrahams, E. Effective Exchange Interactions
for Bad Metals and Implications for Iron-based Superconductors. arXiv:1410.8118.


\bibitem{Goswami.2010}
Goswami, P., Nikolic, P. \& Si, Q.
Superconductivity in Multi-orbital $t$-$J_1$-$J_2$ Model and its Implications for Iron Pnictides. {\it EuroPhys. Lett.} {\bf 91}, 37006 (2010).

\bibitem{Graser.2009}
Graser, S., Maier, T. A., Hirschfeld, P. J. \& Scalapino, D. J.
Near-degeneracy of several pairing channels in multiorbital models for the Fe pnictides.
{\it New J. Phys.} {\bf 11}, 025016 (2009).

\bibitem{Kuroki.2008}
Kuroki, K. {\it et al.},
Unconventional Pairing Originating from the Disconnected Fermi Surfaces of Superconducting
LaFeAsO$_{1-x}$F$_x$.
{\it Phys. Rev. Lett.} {\bf 101}, 087004 (2008).

\bibitem{FWang.2009}
Wang, F., Zhai, H., Ran, Y., Vishwanath, A. \& Lee, D.-H.
Functional Renormalization-Group Study of the Pairing Symmetry and Pairing Mechanism
of the FeAs-Based High-Temperature Superconductor.
{\it Phys. Rev. Lett.} {\bf 102}, 047005 (2009).

\bibitem{Yu.2014b}
Yu, R., Zhu, J.-X., and Si, Q.
Orbital-selective superconductivity, gap anisotropy, and spin resonance excitations in a multiorbital t-J1-J2 model for iron pnictides.
{\it Phys. Rev. B} {\bf 89}, 024509 (2014).

\bibitem{Ge2013}
Ge, Q. {\it et al.},
Anisotropic but Nodeless Superconducting Gap in the Presence of Spin-Density Wave in Iron-Pnictide Superconductor NaFe$_{1-x}$Co$_x$As.
{\it Phys. Rev. X} {\bf 3}, 011020 (2013).

\bibitem{CZhang1}
Zhang, C. {\it et al.}
Double spin resonances and gap anisotropy in superconducting
underdoped NaFe$_{0.985}$Co$_{0.015}$As.
{\it Phys. Rev. Lett.} {\bf 111}, 207002 (2013).

\bibitem{CZhang2}
Zhang, C. {\it et al.}
Neutron spin resonance as a probe of superconducting gap anisotropy in
partially detwinned electron underdoped NaFe$_{0.985}$Co$_{0.015}$As.
{\it Phys. Rev. B} {\bf 91}, 104520 (2015).

\bibitem{Nica.2015} Nica, E., Yu, R. \& Si, Q. Orbital selectivity and emergent superconducting
state from quasi-degenerate $s$- and $d$-wave pairing channels in iron-based superconductors.
arXiv:1505.04170.

\bibitem{Mou11} Mou, D. {\it et al.} Distinct Fermi Surface Topology and
Nodeless Superconducting Gap in a (Tl$_{0.58}$Rb$_{0.42}$)Fe$_{1.72}$Se$_2$
Superconductor. {\it Phys. Rev. Lett.} {\bf 106}, 107001 (2011).

\bibitem{WangDing11} Wang, X. P. {\it et al.} Strong nodeless
pairing on separate electron Fermi surface sheets in (Tl,K)Fe$_{1.78}$Se$_2$ probed by ARPES. {\it Europhys. Lett.} {\bf 93}, 57001 (2011).

\bibitem{WangDing12} Wang, X.-P. {\it et al.} Observation of an isotropic superconducting gap at the Brillouin zone center of Tl$_{0.63}$K$_{0.37}$Fe$_{1.78}$Se$_2$. {\it Europhys. Lett.} {\bf 99}, 67001 (2012).

\bibitem{ParkKeimer11} Park, J. T. {\it et al.} Magnetic Resonant Mode in the Low-Energy
Spin-Excitation Spectrum of Superconducting Rb$_2$Fe$_4$Se$_5$ Single
Crystals. {\it Phys. Rev. Lett.} {\bf 107}, 177005 (2011).

\bibitem{Friemel12} Friemel, G. {\it et al.} Reciprocal-space structure and dispersion
of the magnetic resonant mode in the superconducting phase
of Rb$_x$Fe$_{2-y}$Se$_2$ single crystals. {\it Phys. Rev. B} {\bf 85}, 140511(R) (2012).

\bibitem{HDing.2015} Miao, H. {\it et al.} Observation of strong electron pairing on bands without Fermi surfaces in LiFe$_{1-x}$Co$_x$As. {\it Nat. Communu.} {\bf 6}, 6056 (2015).

\bibitem{Feng.2015} Niu, X. H. {\it et al.} Identification of prototypical Brinkman-Rice Mott physics in a class of iron chalcogenides superconductors. arXiv:1506.04018.

\bibitem{Gooch09}
M. Gooch, B. Lv, B. Lorenz, A. M. Guloy \& C.-W. Chu,
Evidence of quantum criticality in the phase diagram of K$_x$Sr$_{1-x}$Fe$_2$As$_2$
from measurements of transport and thermoelectricity.
Phys. Rev. B {\bf 79}, 104504 (2009).

\bibitem{Ning10} Ning, F. L.
{\it et al.}
Contrasting Spin Dynamics between Underdoped and Overdoped Ba(Fe$_{1?x}$Co$_x$)$_2$As$_2$.
{\it Phys. Rev. Lett.} {\bf 104}, 037001 (2010).

\bibitem{yoshizawa}
Yoshizawa, M. {\it et al.} Structural Quantum Criticality and Superconductivity in Iron-Based Superconductor Ba(Fe$_{1-x}$Co$_x$)$_2$As$_2$.
{\it J. Phys. Soc. Jpn.} {\bf 81}, 024604 (2012).

\bibitem{nni08} Ni, N.
{\it et al.} Effects of Co substitution on thermodynamic and transport properties and anisotropic
$H_{c2}$ in Ba(Fe$_{1-x}$Co$_x$)$_2$As$_2$ single crystals.
 {\it Phys. Rev. B} {\bf 78}, 214515 (2008).

 \bibitem{jhchu09} Chu, J.-H., Analytis, J. G., Kucharczyk, C., and Fisher, I. R.
 Determination of the phase diagram of the electron-doped superconductor
 Ba(Fe$_{1-x}$Co$_x$)$_2$As$_2$.  {\it Phys. Rev. B} {\bf 79}, 014506 (2009).

 \bibitem{clester} Lester, C.
 {\it et al.} Neutron scattering study of the interplay between structure and magnetism
 in Ba(Fe$_{1-x}$Co$x$)$_2$As$_2$.
{\it Phys. Rev. B} {\bf 79}, 144523 (2009).

\bibitem{nandi}
Nandi, S.
{\it et al.} Anomalous Suppression of the Orthorhombic Lattice Distortion in Superconducting Ba(Fe$_{1-x}$Co$_x$)$_2$As$_2$ Single Crystals.
{\it Phys. Rev. Lett.} {\bf 104}, 057006 (2010).

\bibitem{Zheng.2013} Zhou, R. {\it et al.} Quantum Criticality in Electron-doped BaFe$_{2-x}$Ni$_x$As$_2$. {\it Nat. Commun.} {\bf 4}, 2265 (2013).


\bibitem{Lu-prb2014} Lu, X. {\it et al.} Short-range cluster spin glass near optimal superconductivity in BaFe$_{2-x}$Ni$_x$As$_2$. {\it Phys. Rev. B} {\bf 90}, 024509 (2014).

\bibitem {MYi-S-doping2015} Yi, M. {\it et al.}
Electron Correlation-Tuned Superconductivity in Rb$_{0.8}$Fe$_2$(Se$_{1-z}$S$_z$)$_2$.
arXiv:1505.06636.

\bibitem{Reid12} Reid, J.-Ph. {\it et al.} Universal Heat Conduction in the Iron Arsenide Superconductor KFe$_2$As$_2$: Evidence of a $d$-Wave State. {\it Phys. Rev. Lett.} {\bf 109}, 087001 (2012).

\bibitem{Okazaki12} Okazaki, K. {\it et al.} Octet-Line Node Structure of Superconducting Order Parameter in KFe$_2$As$_2$. {\it Science} {\bf 337}, 1314 (2012).

\bibitem{HongLi13} Hong, X. C. {\it et al.} Nodal gap in iron-based superconductor CsFe$_2$As$_2$ probed by quasiparticle heat transport. {\it Phys. Rev. B} {\bf 87}, 144502 (2013).

\bibitem{ZhangLi15} Zhang, Z. {\it et al.} Heat transport in RbFe$_2$As$_2$ single crystals:
Evidence for nodal superconducting gap. {\it Phys. Rev. B} {\bf 91}, 024502 (2015).

\bibitem{Hardy13} Hardy, F. {\it et al.} Evidence of Strong Correlations and
Coherence-Incoherence Crossover in the Iron Pnictide Superconductor KFe$_2$As$_2$.
{\it Phys. Rev. Lett.} {\bf 111}, 027002 (2013).

\bibitem{WangChen13} Wang, A. F. {\it et al.} Calorimetric study of single-crystal
CsFe$_2$As$_2$. {\it Phys. Rev. B} {\bf 87}, 214509 (2013).

\bibitem{Grube2015}
Eilers, F. {\it et al.} Quantum criticality in $A$Fe2As2 with $A$ = K, Rb, and Cs suppresses superconductivity.
arXiv:1510.01857.

\bibitem{Yi_NaFeAs12}
Yi, M. {\it et al.}
Electronic reconstruction through the structural and
magnetic transitions in detwinned NaFeAs.
\textit{New J. Phys.} \textbf{14}, 073019 (2012).

\bibitem{bohmer_shear_2014}
B\"{o}hmer, A. E. {\it et al.}
Nematic Susceptibility of Hole-Doped and Electron-Doped BaFe$_2$As$_2$ Iron-Based
Superconductors from Shear Modulus Measurements.
{\it Phys. Rev. Lett.} {\bf 112}, 047001 (2015).

\bibitem{thorsmolle2014}
Thorsmolle, V. K. {\it et al.}
Critical Charge Fluctuations in Iron Pnictide Superconductors.
arXiv:1410.06116.

\bibitem{kretzschmar2015}
Kretzschmar, F. {\it et al.}
Nematic fuctuations and the magneto-structural phase transition in Ba(Fe$_{1-x}$Co$_x$)$_2$As$_2$.
arXiv:1507.06116.

\bibitem{Song2015}
Song, Y. {\it et al.}
Energy dependence of the spin excitation anisotropy in uniaxial-strained BaFe$_{1.9}$Ni$_{0.1}$As$_{2}$.
{\it Phys. Rev. B} {\bf 92}, 180504 (R) (2015).

\bibitem{Anisimov2002}
Anisimov, V. I. {\it et al.}
Orbital-selective Mott-insulator transition in Ca$_{2-x}$Sr$_x$RuO$_4$.
{\it Eur.~Phys.~J.~B} {\bf 25}, 191-201 (2002).

\bibitem{Neupane2009}
Neupane, M. {\it et al.}
Observation of a Novel Orbital Selective Mott Transition in Ca$_{1.8}$Sr$_{0.2}$RuO$_4$.
{\it Phys. Rev. Lett.} {\bf 103}, 097001 (2009).

\bibitem{HosonoKuroki.2015} Hosono, H. \& Kuroki, K. Iron-based superconductors: Current status of materials and pairing Mechanism. {\it Physica C} {\bf 514}, 399 (2015).

\end{thebibliography}
\end{document}